\documentclass[review,number,sort&compress]{elsarticle}
\input{ds.def}
\usepackage{subcaption}
\DeclareMathOperator{\erf}{erf}
\DeclareMathOperator{\erfc}{erfc}
\newcommand\figurewidth{\columnwidth}
\newcommand{\arthreenine}{\mbox{$^{39}$Ar }}

\address[APC]{APC, Universit\'e Paris Diderot, CNRS/IN2P3, CEA/Irfu, Obs de Paris, USPC, Paris 75205, France}
\address[AQLNGS]{INFN Laboratori Nazionali del Gran Sasso, Assergi (AQ) 67100, Italy}
\address[AQGSSI]{Gran Sasso Science Institute, L'Aquila 67100, Italy}
\address[Augustana]{Physics Department, Augustana University, Sioux Falls, SD 57197, USA}
\address[Belgorod]{Radiation Physics Laboratory, Belgorod National Research University, Belgorod 308007, Russia}
\address[BHSU]{School of Natural Sciences, Black Hills State University, Spearfish, SD 57799, USA}
\address[CAINFN]{INFN Cagliari, Cagliari 09042, Italy}
\address[CAUniPHY]{Physics Department, Universit\`a degli Studi di Cagliari, Cagliari 09042, Italy}
\address[Campinas]{Physics Institute, Universidade Estadual de Campinas, Campinas 13083, Brazil}
\address[CTLNS]{INFN Laboratori Nazionali del Sud, Catania 95123, Italy}
\address[ENUniCEE]{Engineering and Architecture Faculty, Universit\`a di Enna Kore, Enna 94100, Italy}
\address[FNAL]{Fermi National Accelerator Laboratory, Batavia, IL 60510, USA}
\address[GEUni]{Physics Department, Universit\`a degli Studi di Genova, Genova 16146, Italy}
\address[GEINFN]{INFN Genova, Genova 16146, Italy}
\address[Hawaii]{Department of Physics and Astronomy, University of Hawai'i, Honolulu, HI 96822, USA}
\address[Houston]{Department of Physics, University of Houston, Houston, TX 77204, USA}
\address[IHEP]{Institute of High Energy Physics, Beijing 100049, China}
\address[INSTM]{Interuniversity Consortium for Science and Technology of Materials, Firenze 50121, Italy}
\address[JINR]{Joint Institute for Nuclear Research, Dubna 141980, Russia}
\address[Krakow]{M. Smoluchowski Institute of Physics, Jagiellonian University, 30-348 Krakow , Poland}
\address[Kurchatov]{National Research Centre Kurchatov Institute, Moscow 123182, Russia}
\address[LPNHE]{LPNHE, Universit\'e Pierre et Marie Curie, CNRS/IN2P3, Sorbonne Universit\'es, Paris 75252, France}
\address[MEPhI]{National Research Nuclear University MEPhI, Moscow 115409, Russia}
\address[MIINFN]{INFN Milano, Milano 20133, Italy}
\address[MIUni]{Physics Department, Universit\`a degli Studi di Milano, Milano 20133, Italy}
\address[MSU]{Skobeltsyn Institute of Nuclear Physics, Lomonosov Moscow State University, Moscow 119991, Russia}
\address[NAINFN]{INFN Napoli, Napoli 80126, Italy}
\address[NAUniPHY]{Physics Department, Universit\`a degli Studi ``Federico II'' di Napoli, Napoli 80126, Italy}
\address[Petersburg]{Saint Petersburg Nuclear Physics Institute, Gatchina 188350, Russia}
\address[PGUniCBB]{Chemistry, Biology and Biotechnology Department, Universit\`a degli Studi di Perugia, Perugia 06123, Italy}
\address[PGINFN]{INFN Perugia, Perugia 06123, Italy}
\address[PIINFN]{INFN Pisa, Pisa 56127, Italy}
\address[PIUniPHY]{Physics Department, Universit\`a degli Studi di Pisa, Pisa 56127, Italy}
\address[PNNL]{Pacific Northwest National Laboratory, Richland, WA 99352, USA}
\address[Princeton]{Physics Department, Princeton University, Princeton, NJ 08544, USA}
\address[RMTreINFN]{INFN Roma Tre, Roma 00146, Italy}
\address[RMTreUni]{Mathematics and Physics Department, Universit\`a degli Studi Roma Tre, Roma 00146, Italy}
\address[RMUnoINFN]{INFN Sezione di Roma, Roma 00185, Italy}
\address[RMUnoUni]{Physics Department, Sapienza Universit\`a di Roma, Roma 00185, Italy}
\address[SSUniCHP]{Chemistry and Pharmacy Department, Universit\`a degli Studi di Sassari, Sassari 07100, Italy}
\address[Temple]{Physics Department, Temple University, Philadelphia, PA 19122, USA}
\address[UCDavis]{Department of Physics, University of California, Davis, CA 95616, USA}
\address[UCLA]{Physics and Astronomy Department, University of California, Los Angeles, CA 90095, USA}
\address[UMass]{Amherst Center for Fundamental Interactions and Physics Department, University of Massachusetts, Amherst, MA 01003, USA}
\address[USP]{Instituto de F\'isica, Universidade de S\~ao Paulo, S\~ao Paulo 05508-090, Brazil}
\address[VTech]{Virginia Tech, Blacksburg, VA 24061, USA}

\makeatletter
\def\l@subsubsection#1#2{}
\renewcommand\@dotsep{10000}
\makeatother

\begin{document}

%\begin{frontmatter}

  %--------------- JOURNAL ----------------%
 % \journal{Nucl. Instr. \& Meth. A}

  %------------ ARTICLE TITLE ------------%
  \title{Electroluminescence pulse shape and electron diffusion in liquid argon measured in a dual-phase TPC}

  %--------------- AUTHORS ----------------%
  %\author[ucla]{A.~Fan\corref{corr}}
  %\ead{aldenf@physics.ucla.edu}
  %\cortext[corr]{Corresponding author}
  
  \author[Houston]{P.~Agnes}
\author[USP]{I.~F.~M.~Albuquerque}
\author[PNNL]{T.~Alexander}
\author[Augustana]{A.~K.~Alton}
\author[PNNL]{D.~M.~Asner}\fntext[fn1]{Now at Brookhaven National Lab}
\author[USP]{M.~P.~Ave}
\author[PNNL]{H.~O.~Back}
\author[FNAL]{B.~Baldin}
\author[PIINFN,PIUniPHY]{G.~Batignani}
\author[FNAL]{K.~Biery}
\author[RMUnoINFN]{V.~Bocci}
\author[AQLNGS]{G.~Bonfini}
\author[CAINFN]{W.~Bonivento}
\author[AQGSSI,AQLNGS]{M.~Bossa}
\author[GEUni,GEINFN]{B.~Bottino}
% ...

  %--------------- ADDRESSES ----------------%
  %\address[ucla]{Physics and Astronomy Department, University of California Los Angeles, Los Angeles, CA 90064}
  %\address[label]{Address}

  %--------------- ABSTRACT ----------------%
  \begin{abstract}
    We report the measurement of the longitudinal diffusion constant in liquid argon with the \mbox{DarkSide-50} dual-phase time projection chamber. The measurement is performed at drift electric fields of \SI{100}{V/cm}, \SI{150}{V/cm}, and \SI{200}{V/cm} using high statistics \arthreenine\ decays from atmospheric argon. We derive an expression to describe the pulse shape of the electroluminescence signal (S2) in dual-phase TPCs. The derived S2 pulse shape is fit to events from the uppermost portion of the TPC in order to characterize the radial dependence of the signal. The results are provided as inputs to the measurement of the longitudinal diffusion constant $D_L$, which we find to be \SI{4.12 \pm 0.09 }{cm^2/s} for a selection of \SI{140}{keV} electron recoil events in \SI{200}{V/cm} drift field and \SI{2.8}{kV/cm} extraction field. To study the systematics of our measurement we examine datasets of varying event energy, field strength, and detector volume yielding a weighted average value for the diffusion constant of \SI{4.09 \pm 0.12 }{cm^2/s}. The measured longitudinal diffusion constant is observed to have an energy dependence, and within the studied energy range the result is systematically lower than other results in the literature. 
  \end{abstract}

  %--------------- KEYWORDS ----------------%
 % \pacs{Electron Diffusion constant, Liquid argon, Time projection chamber}
  \maketitle

%\end{frontmatter}

\section{Introduction}

DarkSide-50 is the current phase of the DarkSide dark matter search program, operating underground at the Laboratori Nazionali del Gran Sasso in Italy. The detector is a dual-phase (liquid-gas) argon Time Projection Chamber (TPC), designed for the direct detection of Weakly Interacting Massive Particles (WIMPs), and housed within a veto system of liquid scintillator and water Cherenkov detectors. DarkSide-50 has produced WIMP search results using both  atmospheric argon (AAr)~\cite{Agnes:2015gu}
and underground argon (UAr)~\cite{Agnes:2016fz}, which is substantially reduced in \arthreenine\ activity. 

The TPC is filled with liquid argon (LAr) with a thin layer of gaseous argon (GAr) at the top. Ionizing radiation in the active volume of the LAr TPC deposits energy in the form of excitation and ionization. This process leads to the formation of excited dimers $\mathrm{Ar}_{2}^{*}$ whose de-excitation produces prompt scintillation light called S1. The liquid volume is subjected to a uniform drift electric field, causing ionization electrons that escape recombination to drift to the surface of the LAr. The drifted electrons are extracted into and drifted across the GAr by a stronger extraction field, producing electroluminescence light called S2.  The S2 signal provides 3D position information: longitudinal position is given by the drift time of the electrons and transverse position is given by the light distribution over the photomultiplier tubes (PMTs). Pulse shape discrimination on S1 and the ratio S2/S1 allows discrimination between nuclear recoils and electron recoils in the LAr. 

\begin{figure}
\centering
 \includegraphics[width=\figurewidth]{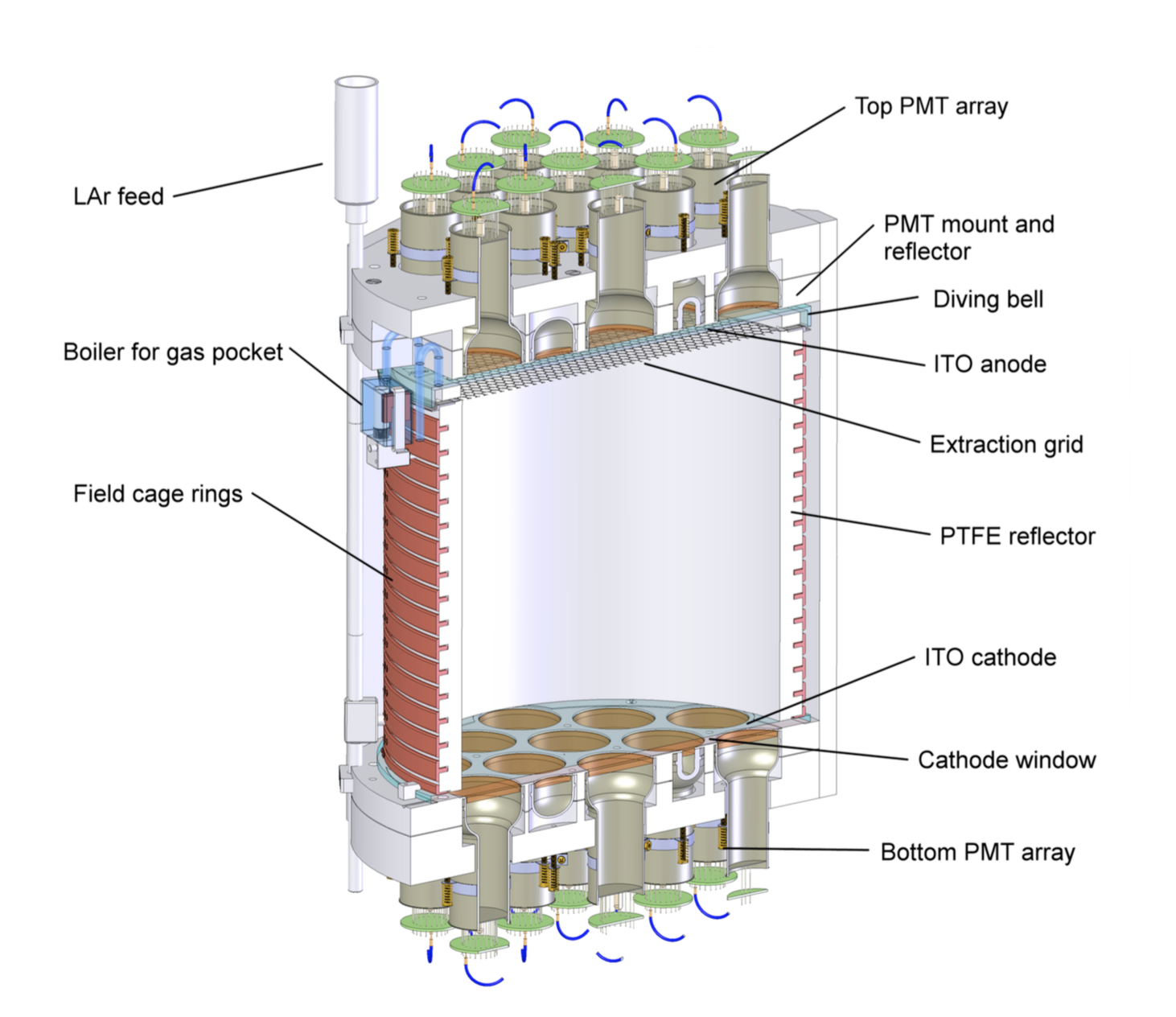}
  \caption{An illustration of the Darkside-50 TPC. The height from the bottom fused silica cathode window to the extraction grid is \SI{35.6}{cm} at room temperature. The LAr surface is slightly above the grid.}
\label{fig:tpc}
\end{figure}

The active volume of the LAr TPC is defined by a \SI{35.6}{cm} diameter by \SI{35.6}{cm} height cylinder. The wall is a monolithic piece of PTFE, the bottom surface is defined by a fused silica window, and the top is defined by a stainless steel grid, as shown in Fig.~\ref{fig:tpc}. To be precise, the grid is positioned just below the liquid-gas interface.
All inner PTFE and fused silica surfaces are coated with tetraphenyl butadiene (TPB) to shift the \SI{128}{nm} scintillation light of LAr to \SI{420}{nm} visible light. 
The S1 and S2 signals are detected by two arrays of 19 PMTs at the top and bottom of the TPC with waveform readout at \SI{250}{MHz} sampling rate. The data acquisition is triggered on S1 and records waveforms for \SI{20}{\us} before the trigger and several hundred microseconds after the trigger, long enough to capture the maximum electron drift time in the TPC, which is drift field-dependent. More information on the DarkSide-50 detector and its performance can be found in references \cite{Agnes:2017ec, Agnes:2017ck, Agnes:2017cl, Agnes:2017cz, Agnes:2016fw, Agnes:2016cp}.

In this work, we analyze the time spectrum of the S2 pulse to investigate the longitudinal diffusion of ionization electrons as a function of drift time in the LAr. The majority of the data used in this analysis were taken as part of the dark matter search campaign using AAr~\cite{Agnes:2015gu}, which is dominated by \SI{1}{Bq/kg} of \arthreenine\ activity \cite{Loosli:1983bu,Benetti:2007fg}. 
The dark matter search data were taken with \SI{200}{V/cm} drift electric field and \SI{2.8}{kV/cm} extraction electric field, corresponding to a \SI{4.2}{kV/cm} electroluminescence field in the gas region. The drift speed of electrons in the \LAr~ for this field configuration is \SI{0.93(1)}{mm/\us} \cite{Agnes:2015gu}, with a maximum drift time of \SI{376}{\us}. Data taken with \SI{150}{V/cm} and \SI{100}{V/cm} drift fields and \SI{2.3}{kV/cm} extraction field are used to study the systematic uncertainties of the longitudinal diffusion measurement. 
 
As a cloud of ionization electrons drifts through the liquid the random walk of the thermalized, or nearly thermalized, electrons will cause the cloud to diffuse over time. The diffusion in the longitudinal and transverse directions, relative to the drift direction, need not be the same. In DarkSide-50, we are sensitive to the longitudinal diffusion, which manifests as a smearing of the S2 pulse shape in time. Previous measurements of electron diffusion in liquid argon have been performed in single phase \TPCs
~\cite{Cennini:1994ba,Li:2016dz} 
where the charge is read out directly. 
This work represents the first measurement of electron diffusion using a dual-phase $\LAr$ TPC.

We assume that the initial size $\sigma_0'$ of a cloud of ionization electrons is of the same order as the recoiled electron track (about $\SI{30}{\um}$ root mean square (RMS) based on a G4DS simulation \cite{Agnes:2017cz} of \SI{140}{keV} electron recoils in \LAr), which is small compared to the eventual size due to diffusion.
Then, if the electrons follow a Gaussian distribution with standard deviation, $\sigma_0'$, centered at a point $(x,y,z)=(0,0,0)$ at time $t=0$, their distribution after drifting a time $t_d$ is given by~\cite{Huxley:1974wj}
\begin{equation} 
  \label{eqn:diff_full}
  n(\rho, z, t_d) = \frac{n_0}{2\pi (2 D_T t_d + \sigma_0'^2) \sqrt{2\pi (2 D_L t_d + \sigma_0'^2)} } \exp \left( - \frac{\rho^2}{4D_T t_d + 2 \sigma_0'^2} \right) \exp \left( - \frac{(z-vt_d)^2}{4D_L t_d + 2\sigma_0'^2} \right)
\end{equation}

where $n_0$ is the number of initial ionization electrons, $v_d$ is the drift velocity in the liquid, $D_T$ is the transverse diffusion coefficient, $D_L$ is the longitudinal diffusion coefficient, $\rho^2 = x^2+y^2$, and $z$ is defined parallel to the drift direction. In DarkSide-50, the electron drift lifetime is \SI{> 5}{ms} \cite{Agnes:2015gu}, corresponding to exceptionally low impurity levels. We therefore neglect the loss of free electrons to negative impurities, so that the integral of $n(\rho, z, t_d)$ over space returns the constant $n_0$ for every $t_d$.

From Eqn.~\ref{eqn:diff_full}, we see that the longitudinal profile of the electron cloud is a Gaussian wave which broadens over time:
\begin{equation}
  \label{eqn:diffusion}
  \sigma_L^2 = 2D_L t_d + \sigma_0'^2
\end{equation}
where $\sigma_L$ is the width of the wave. When the width of the wave grows slowly compared to the drift velocity in the liquid, the diffusion of the electron cloud manifests as a simple Gaussian smearing of the S2 pulse shape. The goal of this analysis is to measure $D_L$, which we achieve by evaluating the smearing $\sigma_L$ as a function of drift time $t_d$ for many events. The smearing is extracted by fitting the S2 pulse shape and the drift time comes directly from the reconstruction. 
In Sec.~\ref{sec:s2pulseshape} we derive an analytic form of the S2 pulse shape. In Sec.~\ref{sec:diffusion} we apply the fitting procedure to various data sets to perform the measurement of electron diffusion in liquid argon.

\section{S2 pulse shape measurement}

\label{sec:s2pulseshape}

The analytic expression for the S2 pulse shape is derived from the following model for the production of light in the gas pocket of the TPC. We assume that ionization electrons drift with constant velocity across the gas pocket, producing Ar excimers uniformly along their drift path. The excimers de-excite and produce light according to a two-component exponential~\cite{Amsler:2008jq}, similar to the light production in the liquid.
If all electrons are extracted from the liquid at precisely the same time, then these two effects define the S2 pulse shape. In reality, electrons of a given ionization cloud are extracted from the liquid with a distribution of times, which we model by introducing a Gaussian smearing term $\sigma$, related to the longitudinal $\sigma_L$ in Eqn.~\ref{eqn:diffusion}, as described in Sec.~\ref{sec:gaussian_smearing}.

\subsection{Basic shape}
\label{sec:basic_shape}

What we will refer to as the basic, or idealized, form of the S2 pulse shape assumes that all electrons are extracted out of the liquid at the same time. It is described by a time profile $y(t)$ given by the convolution of a uniform distribution with a two-component exponential:
\begin{equation}
\label{eqn:s2ideal}
y_\text{ideal}(t; \tau_1,\tau_2,p,T) = p \cdot y'_\text{ideal}(t; \tau_1, T) + {} (1-p) \cdot y'_\text{ideal}(t; \tau_2,T)
\end{equation}
where
\begin{equation} \label{eqn:s2ideal_single}
y'_\text{ideal}(t; \tau, T) = \frac{1}{T}\left\{
\begin{array}{ll}
  0                           ,&  \text{if } t<0 \\
  1-e^{-t/\tau}                ,&  \text{if } 0 \leq t \leq T \\
  e^{-(t-T)/\tau} - e^{-t/\tau}  ,&  \text{if } t>T
\end{array}
\right.
\end{equation}
Here, $\tau_1$ and $\tau_2$ are the fast and slow component lifetimes respectively, $p$ is the fast component fraction, and $T$ is the drift time of the electrons across the gas pocket. We assume that all electrons are extracted out of the liquid at $t=0$. The two decay constants are expected to differ from those of the liquid, the fast and slow components in gas being \SI{11}{\ns} and \SI{3.2}{\us} respectively~\cite{Amsler:2008jq}.
An example pulse shape is shown in Fig.~\ref{fig:s2rebinned_a} in black. Notice that $T$ governs the time to the peak of the pulse. The ``kinks'' in the rising and falling edges are due to the drastically different decay times $\tau_1$ and $\tau_2$, their vertical positions are set by $p$, while their horizontal positions are set by the total drift time in the gas.

\subsection{Gaussian smearing}
\label{sec:gaussian_smearing}

There are many reasons that electrons may not be extracted out of the liquid simultaneously.
The primary reason considered in this analysis is that the cloud of electrons is diffuse, with diffusion arising from drift through the liquid. Minor reasons include the initial size of the cloud of ionization electrons and fluctuations in the time for individual electrons to pass through the grid and the surface of the liquid.
To model the diffusion, we incorporate a smearing term into the S2 pulse shape by convolving Eqn.~\ref{eqn:s2ideal} with a Gaussian centered at 0 with width $\sigma$:
\begin{equation}\label{eqn:s2}
y(t; \tau_1, \tau_2, p, T, \sigma) = y_\text{ideal} \ast \operatorname{gaus}(0, \sigma) =p \cdot y'(t; \tau_1, T, \sigma) + (1-p) \cdot y'(t; \tau_2, T, \sigma)
\end{equation}
where 
\begin{equation}\label{eqn:sigma_relation}
  \sigma^2 = \sigma_L^2/v_d^2 = (\sigma_0^2+2D_L t_d)/v_d^2
\end{equation}

\begin{equation}
y'(t; \tau, T, \sigma) = \frac{1}{2T} \left(  y''(t; \tau, \sigma) - y''(t-T; \tau, \sigma)   \right)
\end{equation}

\begin{equation}\label{eqn:yprimeprime}
y''(t; \tau, \sigma) = \erf \left( \frac{t}{\sqrt{2}\sigma} \right) - e^{-t/\tau}e^{\sigma^2/2\tau^2}\erfc \left( \frac{\sigma^2-t\tau}{\sqrt{2}\sigma\tau}  \right)
\end{equation}
$v_d$ is the drift velocity of electrons in LAr and $\sigma_0$ is a drift-time-independent constant accounting for all the minor smearing effects (initial ionization electron cloud size, additional smearing of the S2 pulse shape in the electroluminescence region, and smearing during electron extraction from the liquid surface). This form has a simple intuitive interpretation: It is the ideal shape of Eqn.~\ref{eqn:s2ideal} with the sharp features smoothed out, as shown in Fig.~\ref{fig:s2rebinned_a} in gray. 

To describe an arbitrary S2 pulse we include three additional parameters in the fit function: a time offset $t_0$, a vertical offset $y_0$, and an overall scale factor $A$. The final fit function is of the form:
\begin{equation}
\label{eqn:s2fitform}
y_\text{fit}(t; \tau_1, \tau_2, p, T, \sigma, A, t_0, y_0) = y_0 + A \cdot y(t-t_0; \tau_1, \tau_2, p, T, \sigma)
\end{equation}
where $\sigma$ is the quantity of interest.

\subsection{Fitting S2 pulse shape}

\label{sec:fittingS2}

We perform event-by-event maximum likelihood fits to the S2 signals. The fits are performed on the summed waveform of all 38 PMT channels of the TPC. Before building the sum waveform, the individual channels are first baseline-subtracted to remove the DC offset in the digitizers, scaled by the single photoelectron (PE) mean, and inverted. The sum waveform is down-sampled, combining every 8 samples together to give \SI{32}{ns} sampling. Single PEs have a FWHM of \SI{\sim 10}{ns}, so down-sampling is performed to reduce bin-to-bin correlations and allow the down-sampled waveform to be interpreted as a histogram of PE arrival times. In the absence of down-sampling, the \SI{250}{MHz} waveform resolution is higher than the single PE width, and the histogram bins are highly correlated. With the down-sampling, though the bins are not integer valued, the bin-to-bin correlations are sufficiently reduced that they approximately follow Poisson statistics. 
Our interpretation of waveforms as histograms has been validated by checking that the bin contents at the same time index of the down-sampled waveforms of events with the same S2 pulse height follow Poisson distributions.

\subsection{Goodness-of-fit}
\label{sec:gof}

\begin{figure*}
\centering
\begin{subfigure}{0.49\textwidth}
  \includegraphics[width=\textwidth]{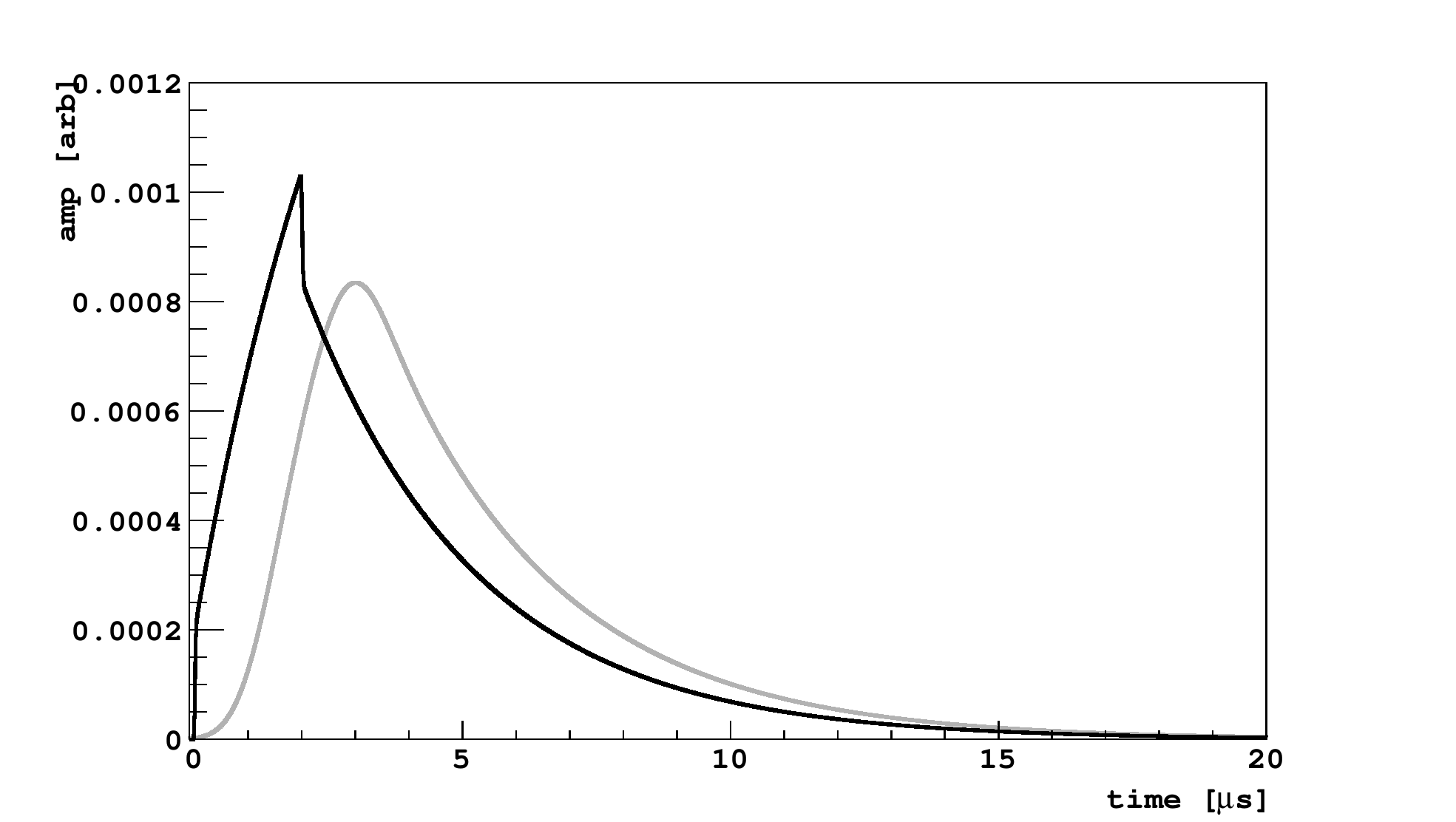}
  \caption{}
  \label{fig:s2rebinned_a}
\end{subfigure}
%\hfill
\begin{subfigure}{0.49\textwidth}
  \includegraphics[width=\textwidth]{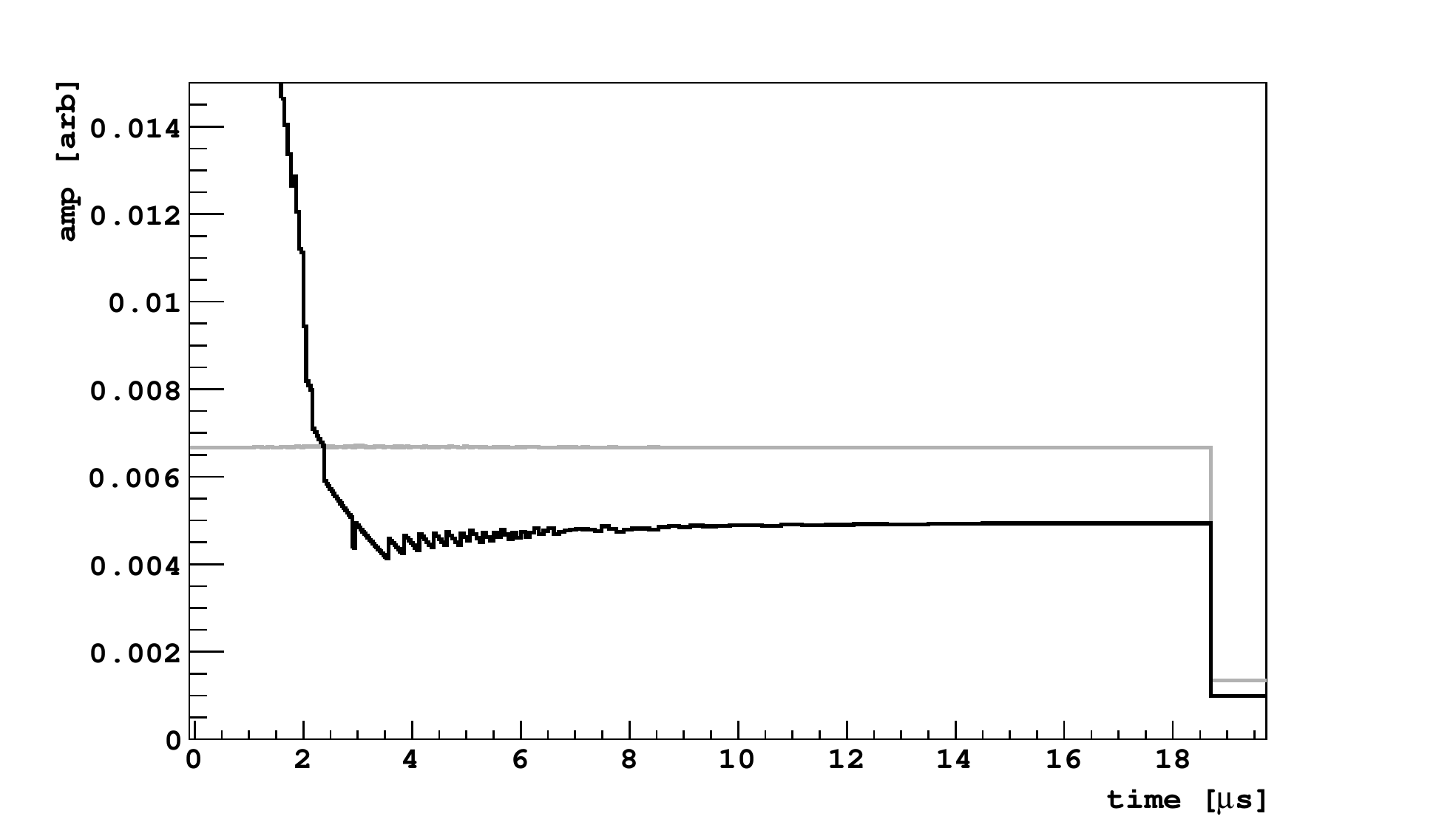}
  \caption{}
  \label{fig:s2rebinned_b}
\end{subfigure}
%\hfill
\begin{subfigure}{0.49\textwidth}
  \includegraphics[width=\textwidth]{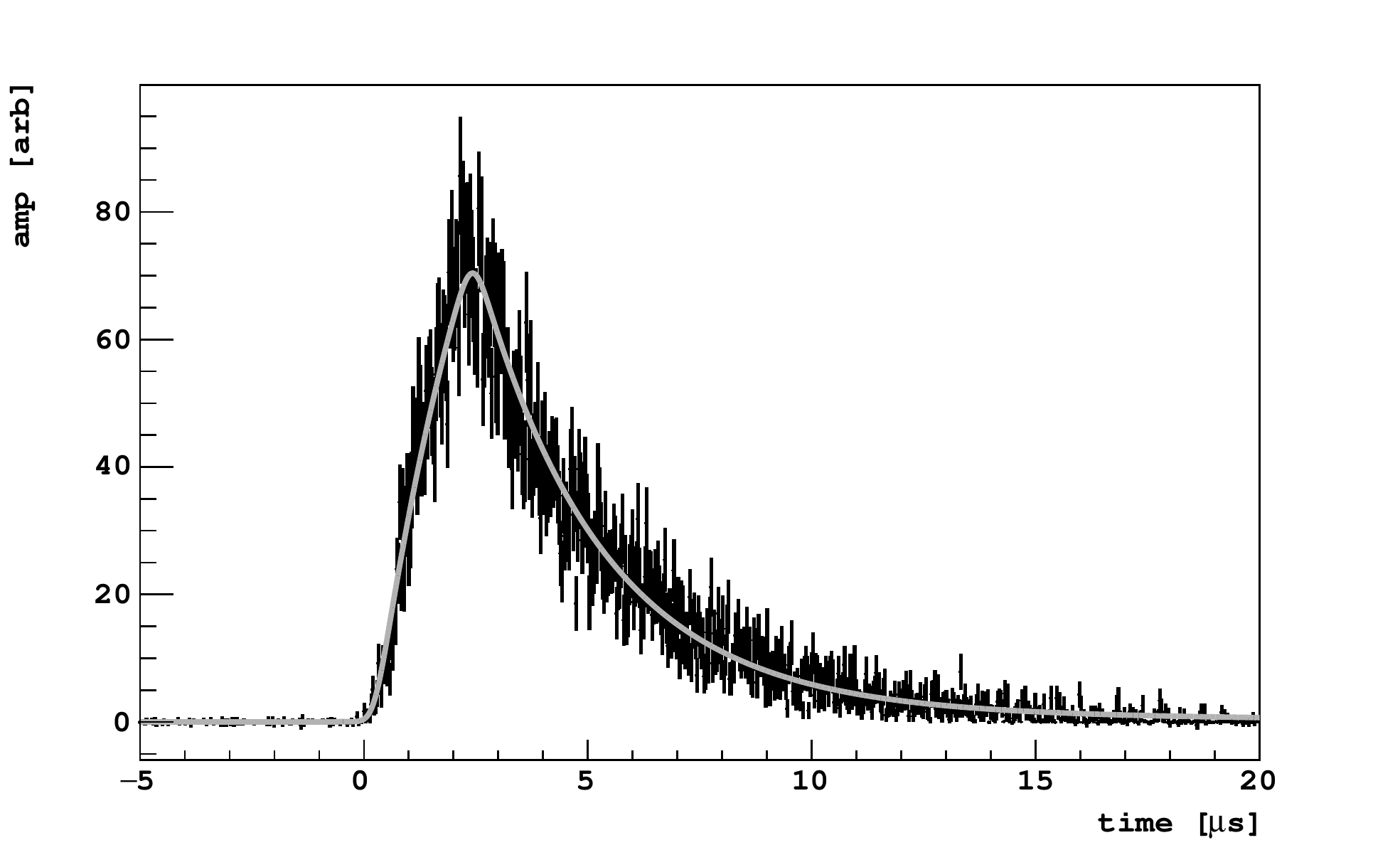}
  \caption{}
  \label{fig:s2rebinned_c}
\end{subfigure}
%\hfill
\begin{subfigure}{0.49\textwidth}
  \includegraphics[width=\textwidth]{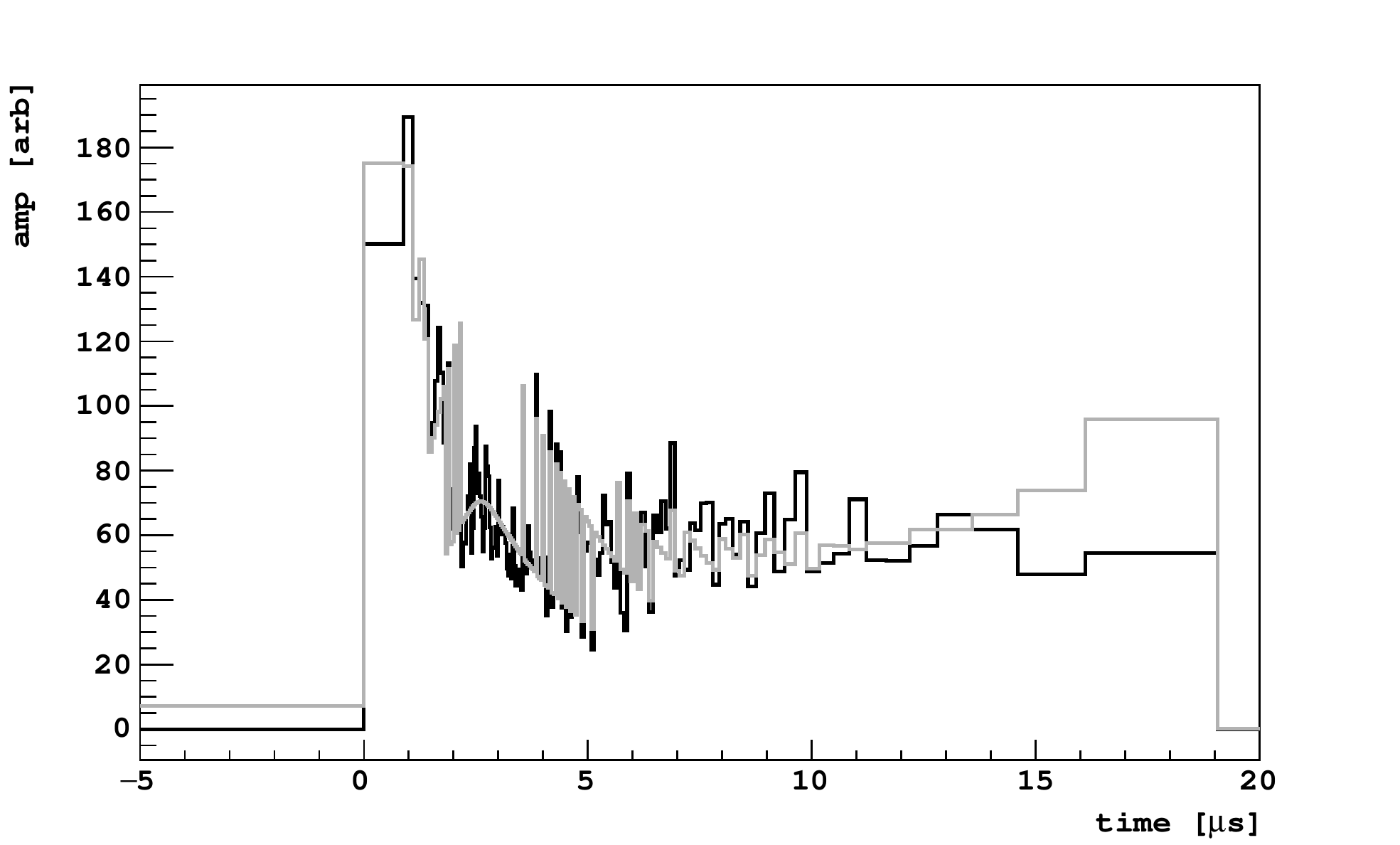}
  \caption{}
  \label{fig:s2rebinned_d}
\end{subfigure}
\caption{(a) Example S2 pulse shape with $\tau_1=$\SI{0.011}{\us}, $\tau_2=$~\SI{3.2}{\us}, $p=0.1$, and $T=$~\SI{1.5}{\us}. \textbf{Black:} Idealized form (no smearing). \textcolor{gray}{\textbf{Gray:}} Includes Gaussian smearing at $\sigma=$~\SI{0.3}{\us}. (b) Re-binned S2 pulse shapes using unequal bin widths, chosen such that the smeared S2 pulse has a flat distribution. The black and gray curves have the same binning. (c)Sample S2 from electronics Monte Carlo (MC) (\textbf{black}) with fitted pulse shape (\textcolor{gray}{gray}). (d) Re-binned versions of (c).}
\label{fig:s2rebinned}
\end{figure*}

To evaluate goodness-of-fit of the S2 pulse shape on the waveforms, we evaluate a $\chi^2$ statistic. However, many of the bins have low (fewer than 5) counts, even after 8 sample re-binning, invalidating a direct $\chi^2$ evaluation. To resolve this issue, we re-bin the waveform again, this time using unequal bin widths. We choose the bin edges so that, for the S2 pulse with moderate smearing ($\sigma = \SI{0.3}{us}$) shown in Fig.~\ref{fig:s2rebinned_a}, each bin has equal counts (Fig.~\ref{fig:s2rebinned_b}). 

The binning is configured so that the minimum bin width is \SI{32}{ns}, and the bin edges are truncated to land on \SI{4}{ns} intervals. For simplicity, we use the same re-binning to evaluate the $\chi^2$ of all events. As the pulse shape varies, the re-binned waveforms will not populate the bins with equal counts, as shown in Fig.~\ref{fig:s2rebinned_b}. However, their shapes will be similar enough and, as we constrained our study to S2 \SI{> e4}{PE}, the bins do not fall below 5 counts. Example waveforms before and after re-binning are shown in Fig.~\ref{fig:s2rebinned_c} and~\ref{fig:s2rebinned_d}.

The $\chi^2$ statistic that we use is the one prescribed by Baker and Cousins~\cite{Baker:1984fn}, reproduced here:
\begin{equation}
  \chi^2 = 2\sum_i y_i - n_i + n_i \ln \left( \frac{n_i}{y_i} \right)
\end{equation}
where the sum is over the bins of the re-binned S2 waveform, $n_i$ is the contents of the $i$th bin, and $y_i$ is the number of PE predicted by the model to be in the $i$th bin.

\subsection{Degeneracy of parameters}

\label{sec:parameter_degeneracy}

\begin{figure}
  \centering
  \includegraphics[width=\figurewidth]{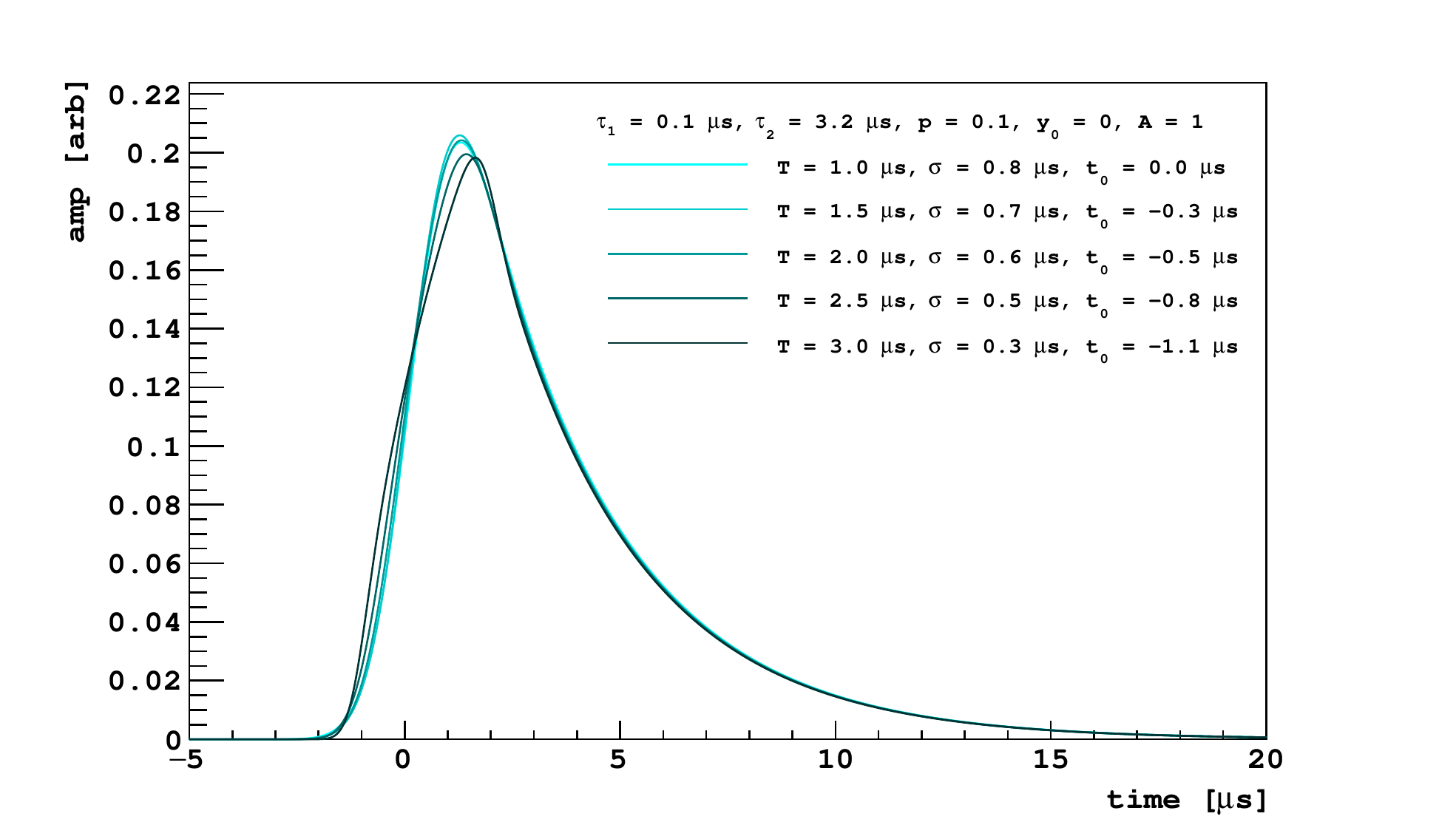}
  \caption{Family of S2 pulses with different values for $\sigma$, $T$, and $t_0$ but nearly identical pulse shape. (For interpretation of the references to color in this figure legend, the reader is referred to the web version of this article.)}
  \label{fig:s2family}
\end{figure}

The form of the S2 pulse shape given in Eqn.~\ref{eqn:s2fitform} has an approximate degeneracy: the same shape can be produced using different combinations of $T$, $t_0$, and $\sigma$.
The degeneracy can be seen visually in Fig.~\ref{fig:s2family}, where five nearly identical pulse shapes are shown using different parameter values.

This degeneracy can result in incorrect parameter estimation if the parameters are all left free in the fit. In order to make a precise estimation of the S2 diffusion parameter $\sigma$, we fix the gas pocket drift time $T$. $T$ is related to gas pocked thickness and electroluminescence field strength, which both exhibit rotational symmetry. $T$ is approximately azimuthally symmetric, and it is sufficient to fix $T$ based on its radial dependence, fitting events with very little diffusion to extract $T(r)$. The relationship between $T$ and $r$ is consistent with a non-uniform electroluminescence field that is strongest at the center of the TPC and gradually weakens towards the edge. There are several possible explanations, including a sagging anode window or a deflecting grid, but we have insufficient information from these results to discriminate between these explanations. 

\subsubsection{Zero diffusion event selection}
\label{sec:zeroDiffEventSelection}
We search for zero-diffusion events in a subset of the high statistics \arthreenine data from the AAr dark matter search dataset. The data used here are at higher energies than those used in the dark matter search analysis, S2 = $(1-5)\times 10^4$ PE, because we require high PE statistics to ensure the quality of the S2 pulse shape fits to individual events. To examine zero diffusion events we select single scatter events from the top of the TPC passing our basic quality cuts requiring that all channels are present in the readout, and that the waveform baselines were found successfully. More precisely, we look for events with S1, S2, and $t_\text{d}$ less than \SI{5}{\us}. 
We fit Eqn.~\ref{eqn:s2fitform} to each event, using the maximum likelihood method described in Sec.~\ref{sec:fittingS2}. 

The 3-parameter degeneracy described in Sec.~\ref{sec:parameter_degeneracy} is not relevant in zero-diffusion events, however, it is broken nonetheless by fixing $\sigma$ to a very small non-zero value to avoid division issues. The slow component term $\tau_2$ can be ``pre-fit'' using the tail of each waveform, where the fast component contribution to the electroluminescence signal is negligible. This is done prior to re-binning, when the fit has more sensitivity to $\tau_2$. We fit a simple exponential decay in the range of \SIrange{9}{20}{\us} of each event, to avoid smearing from the fast decay component and baseline noise, see Fig.~\ref{fig:s2rebinned_c} for reference. This range guarantees that we fit to the tail of the S2 pulse even in events with the highest diffusion, where the peak is farthest from the pulse start. In the full fit we initialize $\tau_2$ to the value from the pre-fit, but leave it free to vary. This improved fitter performance but does not affect the overall results compared to using a global fixed value of $\tau_2$. The amplitude $A$ is initialized to the total area of the waveform. We have now turned an 8 parameter fit into effectively a 5 parameter fit. The remaining parameters are given sensible initial values, as shown in Tab.~\ref{tab:zdfitparams}. Reasonable variation of these initial values did not change the outcome of the fits and  fit results remained within defined parameter limits.
\begin{table}
  \centering
  \begin{tabular}{ c  c }
    %\hline
    Parameter & Initial value \\
    \hline
    $\tau_1$  &  \SI{0.01}{\us} \\
    $\tau_2$ & pre-fit in tail (initialized but not fixed) \\
    $p$ & 0.1 \\
    $T$ & \SI{1.6}{\us} \\
    $\sigma$ & 0.01 (fixed)\\
    $A$ & area of pulse \\
    $t_0$ & 0 \\
    $y_0$ & 0 \\
    %\hline
  \end{tabular}
  \caption{Initial values of fit parameters for zero-diffusion events.}
  \label{tab:zdfitparams}
\end{table}

\subsubsection{Results}

\begin{figure}
\centering
\includegraphics[width=\figurewidth]{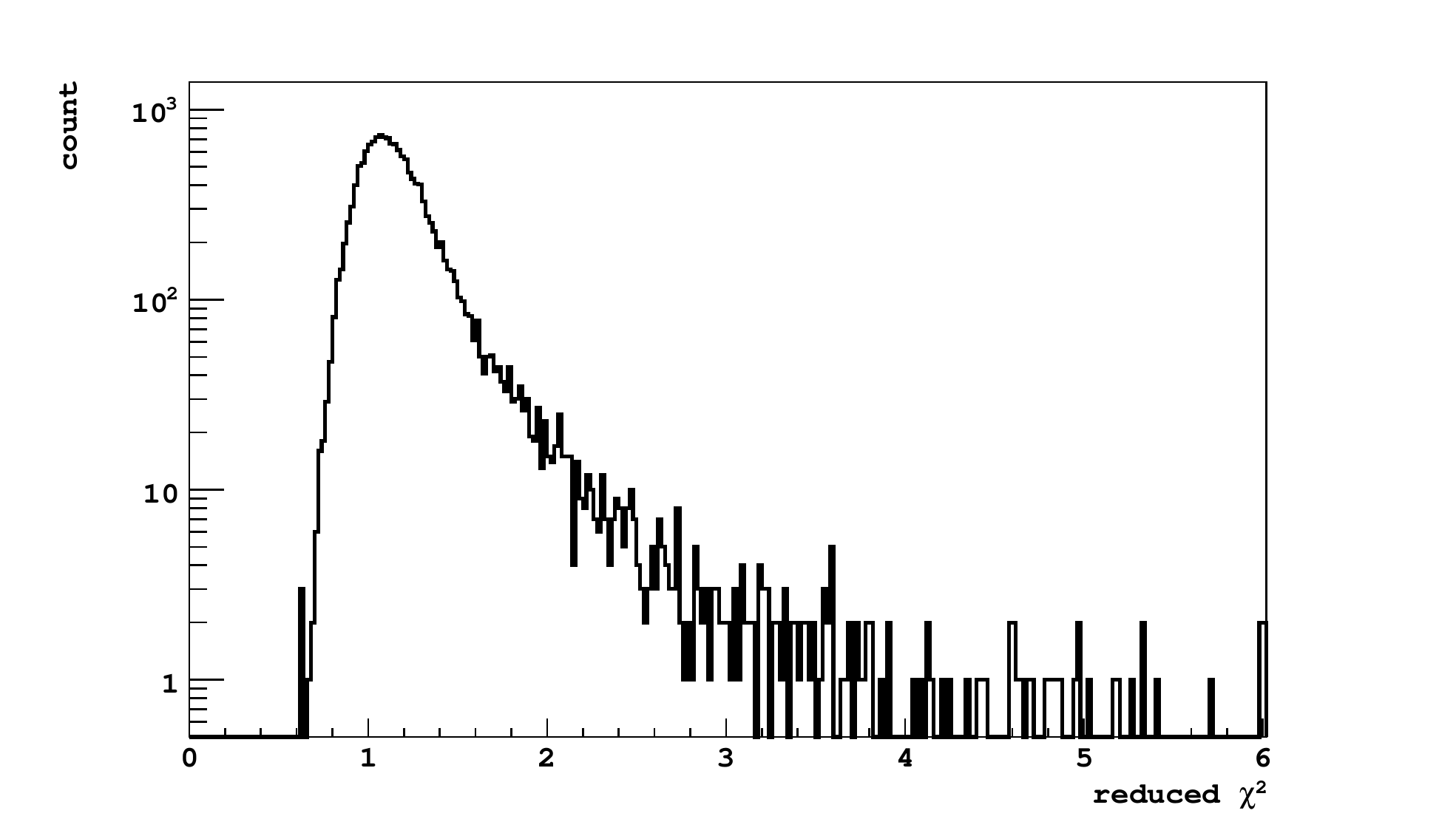}
\caption{Reduced $\chi^2$ of S2 pulse shape fits to \num{1.6e4} zero-diffusion events. }
\label{fig:zdchi2}
\end{figure}

\begin{figure}
\centering
\begin{subfigure}{\figurewidth}
  \includegraphics[width=\textwidth]{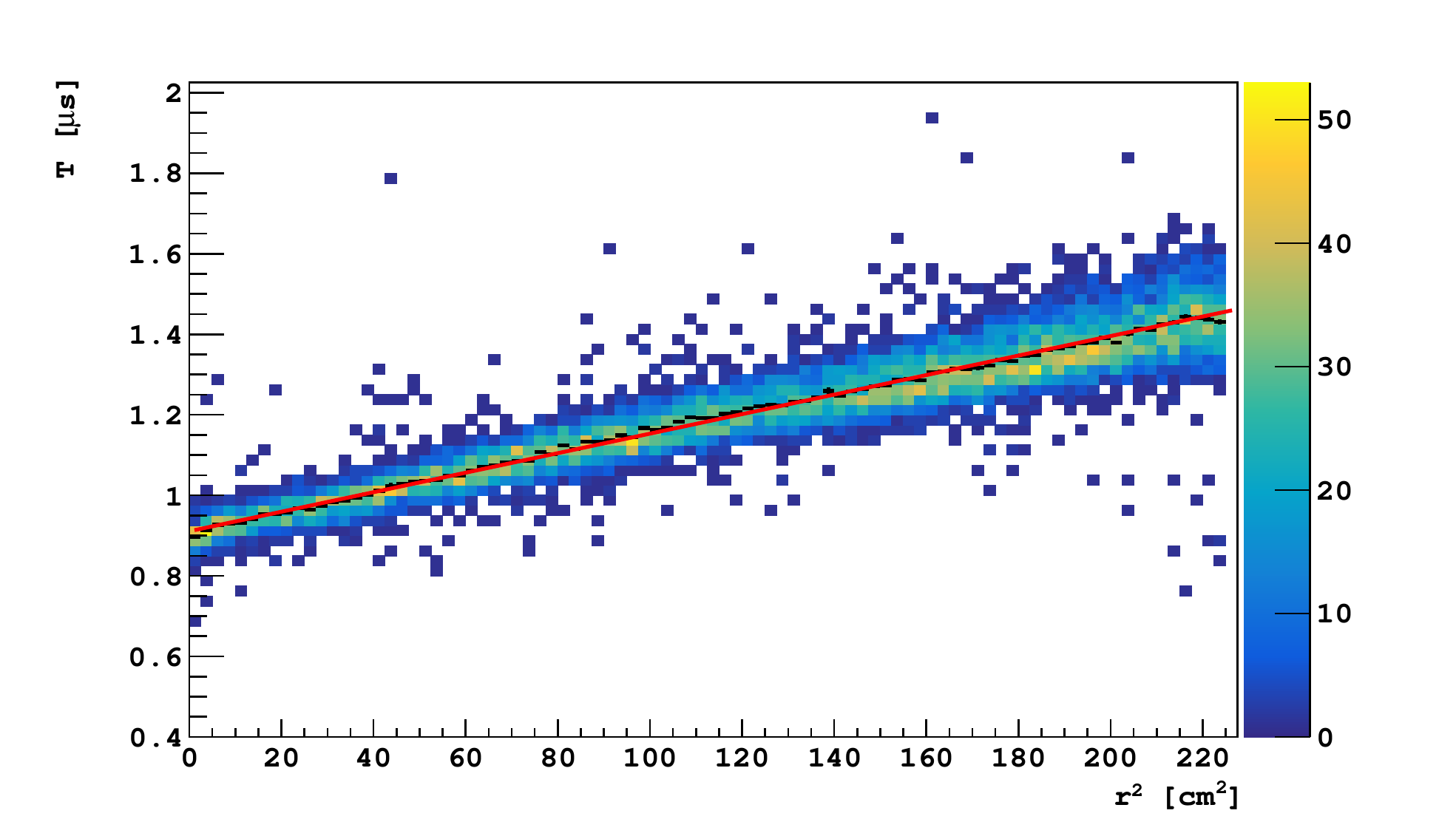}
  \caption{}
  \label{fig:zdfit_a}
\end{subfigure}
\begin{subfigure}{\figurewidth}
  \includegraphics[width=\textwidth]{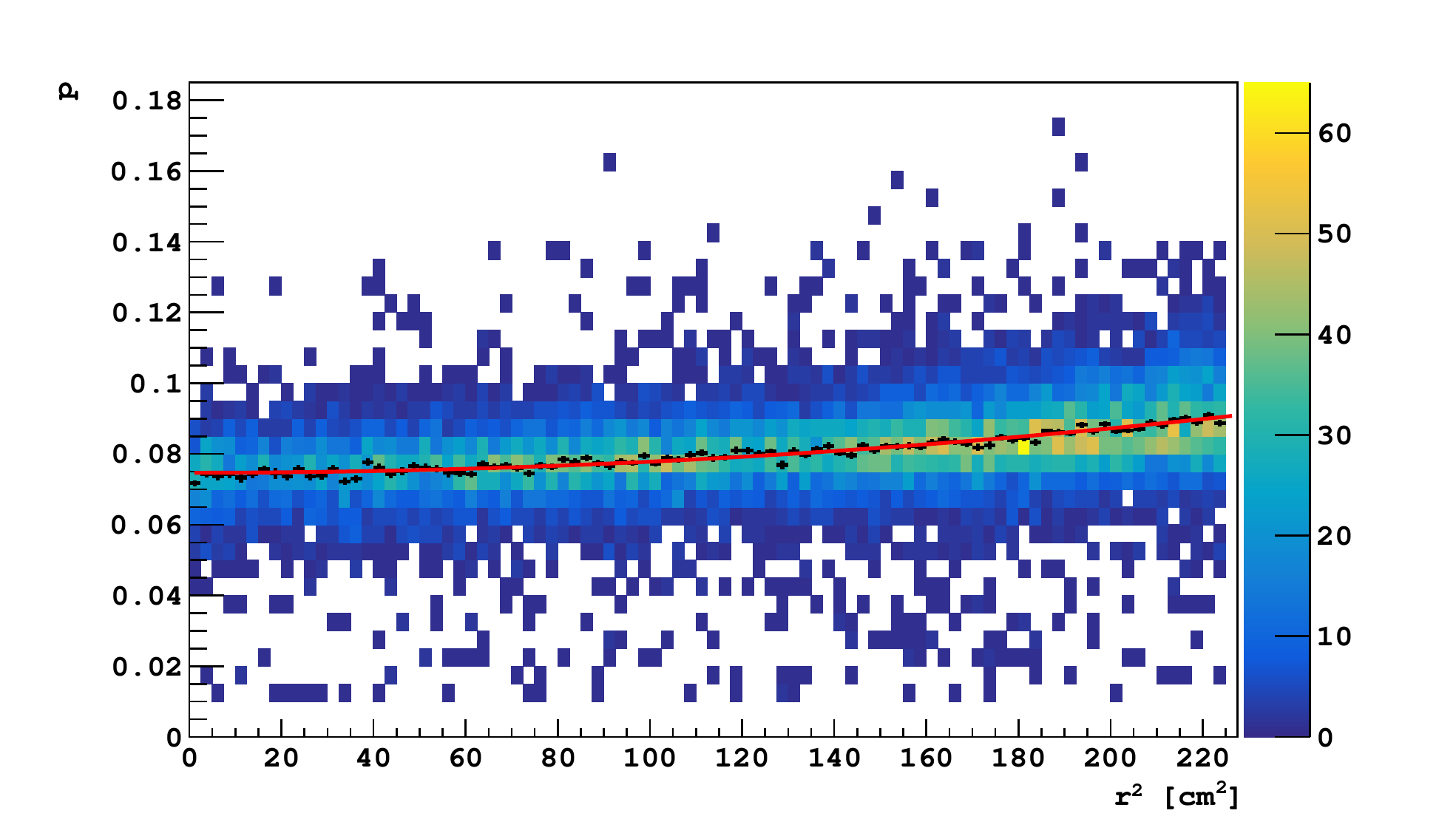}
  \caption{}
  \label{fig:zdfit_b}
\end{subfigure}
\caption{(a) 2D histogram of $T$ vs. $r^2$. The mean values of the bin contents are fit with Eqn.~\ref{eqn:T_R}. The mean values (black points) are under the fit (red curve). (b) 2D histogram of the fast component fraction $p$ vs. $r^2$. The mean values of the bin contents are fit with Eqn.~\ref{eqn:p_R}. The mean values (black points) are under the fit (red curve). (See the web version of this article for color.)}
\label{fig:zdfit}
\end{figure}

We fit the S2 pulse shape to $3.47\times10^4$ zero-diffusion events. The goodness-of-fit is evaluated for each event using the procedure described in Sec.~\ref{sec:gof}. Because we use the same binning and fit function for each fit, the NDF is the same throughout (NDF = 133). The distribution of the reduced $\chi^2$ statistic ($\chi^2_\text{red} = \chi^2/\text{NDF}$) is shown in Fig.~\ref{fig:zdchi2}, zoomed to $\chi^2_\text{red}<6$. 
About \SI{10}{\percent} of events have very poor fits with $\chi^2_\text{red}>1.5$. 
The zero diffusion events exhibit a spectrum of separations between S1 and S2 and there are some events where the signals are so close to each other that they are essentially indistinguishable. 
To avoid these suboptimal events we require $t_0>\text{\SI{-0.1}{\us}}$ and $\chi^2_{\rm red}<1.5$.
After these additional cuts are applied, we plot $T$ from each fit as a function of radial position, as shown in Fig.~\ref{fig:zdfit_a}.

The mean of the $T$ vs. $r^2$ distribution is well fit by a linear function. 
We take the function $T(r)$ to be of the form:
\begin{equation}\label{eqn:T_R}
T(r) = A_T ( 1 + \frac{r^2}{B_T})
\end{equation}
Fitting Eqn.~\ref{eqn:T_R} to the mean of the $T$ vs. $r^2$ distribution, we find 
$A_T = \text{\SI{910.8 \pm 0.8}{\ns}}$ and $B_T=\text{\SI{376 \pm 1}{cm^2}}$. Uncertainties are statistical.

The fits to the zero-diffusion events can also give us information about the fast component fraction $p$ in the gas and the slow component lifetime $\tau_2$. The distribution of $p$ vs. radial position is shown in Fig.~\ref{fig:zdfit_b}. We expect $p$ to depend on the extraction field, as the field strength will affect recombination and therefore the ratio of triplet to singlet states \cite{Chepel:2013ab,Kubota:1978ab}. Since the electroluminescence field varies radially in DarkSide-50, so does $p$.

The relationship between $p$ and $r$ is well fit by a function of the form 
\begin{equation} \label{eqn:p_R}
  p(r) = A_p(1+\frac{r^4}{B_p^2})
\end{equation}
Fitting Eqn.~\ref{eqn:p_R} to the mean of the $p$ vs. $r^2$ distribution we find 
$A_p=(7.47 \pm 0.02)\times 10^{-2}$ and $B_p=\text{\SI{488 \pm 6}{cm^2}}$. Uncertainties are statistical.

\begin{figure}
  \centering
  \includegraphics[width=\figurewidth]{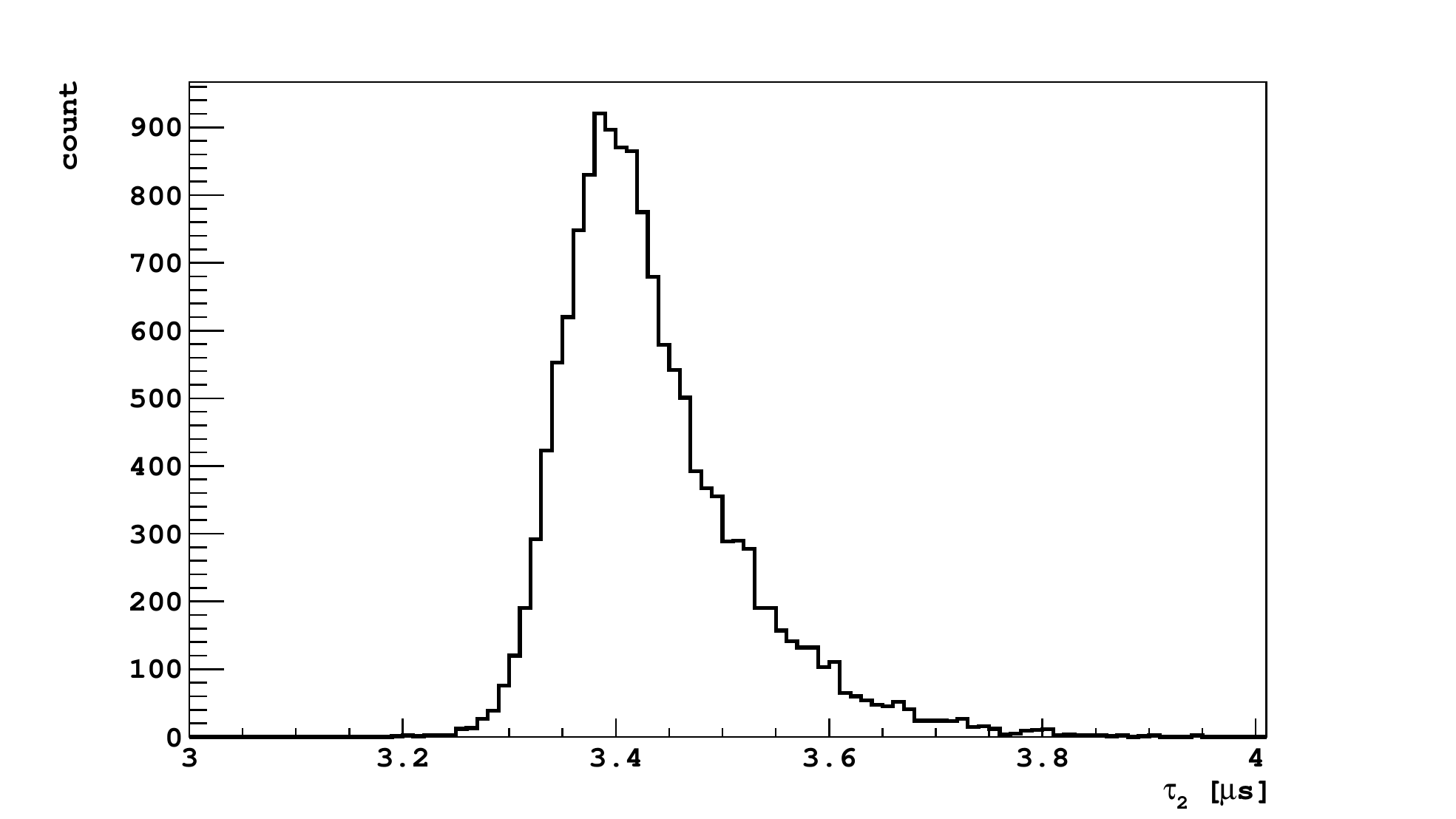}
  \caption{Distribution of the slow component lifetime $\tau_2$, extracted from fits to zero-diffusion events.}
  \label{fig:h_zd_tau2}
\end{figure}

Because of the \SI{32}{ns} binning of the waveforms, we do not have the resolution required to estimate the fast component lifetime $\tau_1$, instead it is fixed to a reasonable, small number in the fits. 
The distribution of $\tau_2$ is shown in Fig.~\ref{fig:h_zd_tau2}.  
The average slow component lifetime is $\tau_2=\text{\SI{3.43}{\us}}$, which agrees well with the previously measured value $\tau_2=\SI{3.2\pm0.3}{\us}$ ~\cite{Keto:1974hj}.

\section{Electron diffusion measurement}
\label{sec:diffusion}

\subsection{Event selection}
\label{sec:diffEventSelection}

The principle data used for this analysis are the abundant \arthreenine decays from AAr data at standard $\SI{200}{V/cm}$ drift field and $\SI{2.8}{kV/cm}$ extraction field, the same data set used in Sec.~\ref{sec:parameter_degeneracy}.
We use additional sets of data to perform cross-checks and systematic uncertainty measurements of the diffusion, including data at different drift and extraction fields.

To perform the measurement of the longitudinal diffusion constant, we use well-reconstructed single scatter \arthreenine events. 
We select events that pass basic quality cuts as discussed in Sec.~\ref{sec:zeroDiffEventSelection}.
We select single scatter events by requiring that the reconstruction software identifies one S1 and one S2 pulse, and that the S1 start time is at the expected trigger time within the acquisition window.
To reduce possible systematics due to variations of $T(r)$ at different $r$, we select events in a narrow $r$ slice: \SIrange{9}{12}{cm}. Finally, we select events with maximum possible PE statistics before the S2 saturates the digitizers: $(4-5)\times \SI{e4}{PE}$. The selected events have a mean S1 of \SI{1000}{PE} with RMS \SI{150}{PE}. The measured S1 light yield in DarkSide-50 is \SI{7.0(3)}{PE/keV} \cite{Agnes:2015gu}, corresponding to a selection of \SI{140(20)}{keV} electron recoils. We repeat the analysis on different $r$ and S2 slices to estimate the systematics.

\subsection{Fitting procedure}
\label{sec:diffFitProcedure}

We perform a fit of the S2 pulse shape on every event that passes the event selection. As in the case of the S2 pulse shape analysis, there are 8 parameters in the fit (Eqn.~\ref{eqn:s2fitform}). Here we describe the choice of initial values for each of those parameters.

\begin{itemize}
  
\item As shown in Fig.~\ref{fig:zdfit}, $T$ varies with transverse position. We fix $T$ on an event-by-event basis, evaluating $T(r)$ as given by Eqn.~\ref{eqn:T_R}.

\item For each event we pre-determine the value of the baseline offset $y_0$ by fitting a flat line to the pre-signal region of \SIrange{-5}{-1}{\us}. The baseline value in the full fit is fixed to the value determined here. 
  
\item The fast component lifetime should be independent of $t_d$. However, $\tau_1$ cannot be well-constrained due to the resolution of our waveforms. Because the fast component gets washed out with any non-negligible amount of smearing, we fix $\tau_1 = \text{\SI{0.01}{\us}}$, close to the value from \cite{Amsler:2008jq}.

\item The slow component lifetime should also be independent of $t_d$, but since it is the principle shape parameter in the long tail of S2, we do not fix it globally. As in the analysis of the zero-diffusion events, we determine $\tau_2$ prior to the full S2 fit by fitting an exponential to the tail of the S2 pulse in the region \SIrange{9}{20}{\us} after the pulse start. The fit function is $y=Ae^{-t/\tau_2}$. The value of $\tau_2$ is initialized to the value from the pre-fit, but left free to vary in the full fit.

\item We do not expect the fast component fraction, $p$, to vary with respect to $t_d$, but it varies with electroluminescence field, and therefore varies with respect to radial position in DarkSide-50. Like $T$, we fix $p$ on an event-by-event basis, evaluating $p(r)$ as given by Eqn.~\ref{eqn:p_R}.

\item The initial value of $\sigma$ is given by the value of diffusion measured in ICARUS, $\sigma_{\rm init} = \sqrt{2 D_L^I t_d}/v$ with $D_L^I=\text{\SI{4.8}{cm^2/s}}$~\cite{Cennini:1994ba}. 

\item The amplitude parameter $A$ is initialized to the total area of the S2 pulse.

\item The time offset parameter $t_0$ is expected to vary with each event: for events with more diffusion, the pulse finding algorithm of the reconstruction will find the pulse start relatively earlier with respect to the pulse peak. We empirically find that $t_0$ varies linearly with $\sigma$: $t_0 = -0.25+3.06\sigma$, which we use to set the initial value of the time offset: $t_{0,\text{init}}=t_0(\sigma_\text{init})$.

\end{itemize}

The initial values of all the fit parameters are summarized in Tab.~\ref{tab:initParams}. 
\begin{table}
  \centering
  \begin{tabular}{ c  c }
    Parameter & Initial Value \\
    \hline
    $\tau_1$  &  \SI{0.01}{\us} (fixed) \\
    $\tau_2$ & pre-fit in tail (initialized but not fixed)\\
    $p$ & $p(R)$ (fixed) \\
    $T$ & $T(R)$ (fixed) \\
    $\sigma$ & $\sqrt{2D_Lt_{\rm d}}/v_{\rm drift}$ \\
    $A$ & area of S2 pulse \\
    $t_0$ & max(-0.25+3.06$\sigma_{\rm init}$, 0) \\
    $y_0$ & pre-fit in pre-signal region  (fixed)\\
  \end{tabular}
  \caption{Initial values of fit parameters.}
  \label{tab:initParams}
\end{table}
Of the original 8 parameters, 4 of them are fixed in the final fit of the S2 pulse shape. The remaining free parameters are $\sigma$, $\tau_2$, $A$, and $t_0$. For each event, we re-define the x-axis such that $t=0$ is at the S2 pulse start time as determined by pulse finding program, and truncate the waveform leaving the \SIrange{-5}{20}{\us} region about the newly defined $t=0$. 
The truncated waveform is down-sampled as discussed in Sec.~\ref{sec:fittingS2}, and fit by the maximum likelihood method.

\subsection{Drift velocity}
\label{sec:driftVelocity}
\begin{figure}
\centering
\includegraphics[width=\figurewidth]{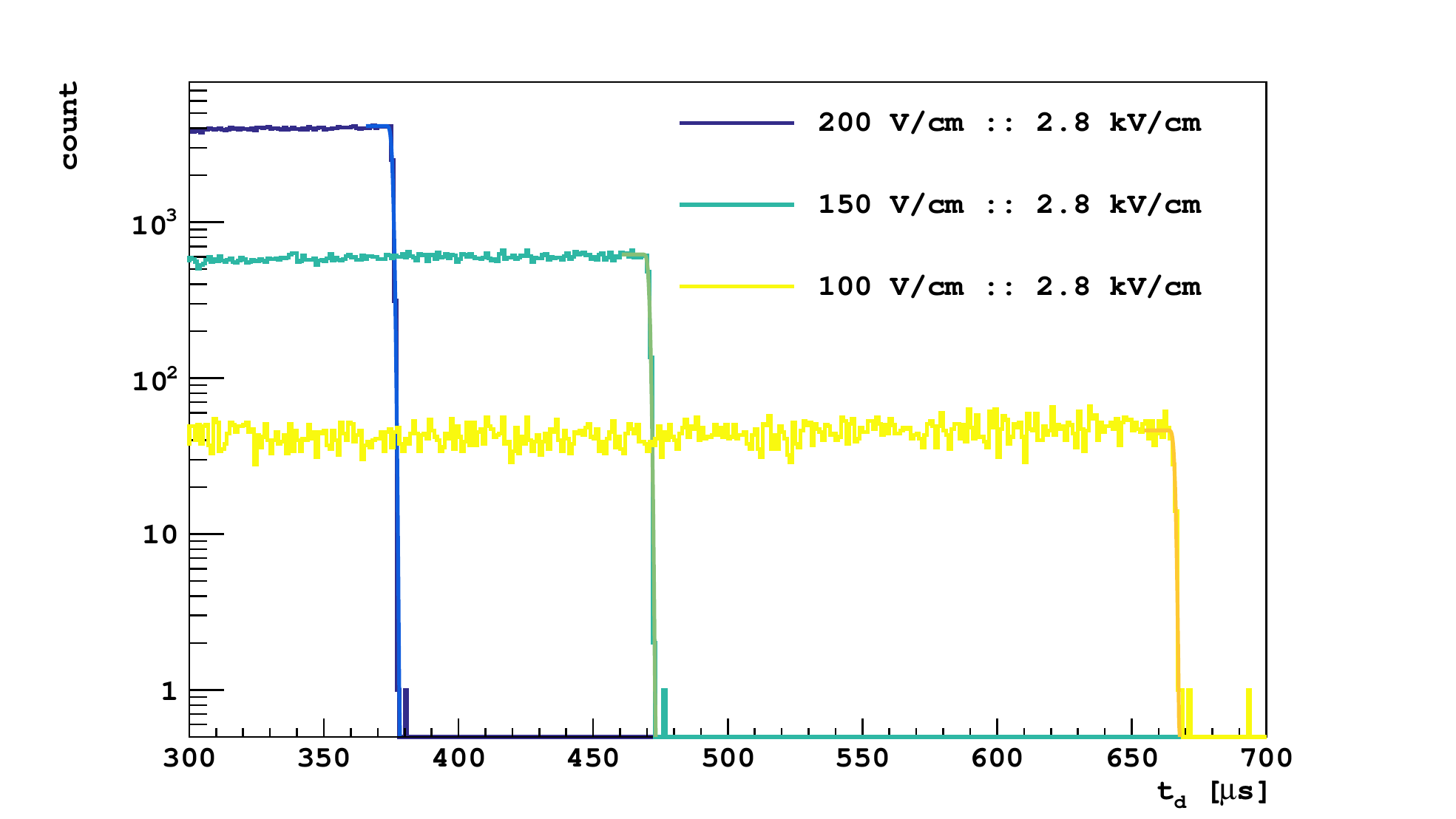}
\caption{Electron drift time distributions under different drift fields. The maximum drift time is defined as the half maximum position to the right of each plateau. Precise values are obtained from a sigmoidal fit utilizing a complementary error function, shown as curves on the right edge of each histogram. (For interpretation of the references to color in this figure legend, the reader is referred to the web version of this article.)}
\label{fig:maxtDrift}
\end{figure}

The electron drift velocity $v_d$ and mobility $\mu$ in LAr under different drift fields are calculated from the maximum drift time, as shown in Fig.~\ref{fig:maxtDrift}, and the height of the TPC drift region. 
The drift time $t_{d}$ is defined as the difference between the start times identified by the reconstruction algorithm for S2 and S1, plus the parameter $t_{0}$ from the fit. The addition of the time offset parameter corrects for the fact that diffusion of the S2 pulse will cause the reconstruction algorithm to identify the S2 start time relatively earlier than for a pulse with zero diffusion. In fact, $t_{0}$ is generally negative.
The height of the TPC region is measured to be $\SI{35.56\pm0.05}{cm}$ at room temperature. The PTFE will contract $\SI{2\pm0.5}{\percent}$ at the operating temperature of $\SI{89.2\pm0.1}{K}$. This contraction, determined through measurements of the DarkSide-50 TPC, is in agreement with \cite{Kirby:1956ce}. $200/150/100$ \si{V/cm} are named referring to the warm Teflon height, but the appropriate height is used in our actual calculations, resulting in slightly higher field values. Uncertainty from field non-uniformity near the grid and the time electrons drift in LAr above the grid are also considered. Field non-uniformity contributes uncertainty to the field strength and therefore the mobility, as we can only measure the voltage on the electrodes. The values shown in Tab.~\ref{tab:vdrift} agree with \cite{Cennini:1994ba} and \cite{Li:2016dz}.

\begin{table*}
  \centering
  \begin{tabular}{c c c c}
    Drift field [\si{V/cm}] & Corrected drift field [\si{V/cm}]  &  $v_d$ [\si{mm/\us}] & $\mu$ [\si{cm^2/Vs}]\\
    \hline
    100 & $101.8\pm0.7$ & $0.524\pm0.004$ & $514\pm5$\\
    150 & $152.7\pm1.0$ & $0.742\pm0.005$ & $485\pm5$\\
    200 & $203.6\pm1.4$ & $0.930\pm0.007$ & $456\pm5$
  \end{tabular}
	\caption{Electron drift velocity and mobility in LAr for different drift fields in DarkSide-50 at \SI{89.2\pm0.1}{K}. Numbers are calculated using the maximum drift time and the height of TPC drift region. The effect of field non-uniformity and PTFE shrinkage are considered in the calculation. The corrected drift field values take PTFE shrinkage into account.}
  \label{tab:vdrift}
\end{table*}

\subsection{Results}

\begin{figure}
\centering
    \includegraphics[width=\figurewidth]{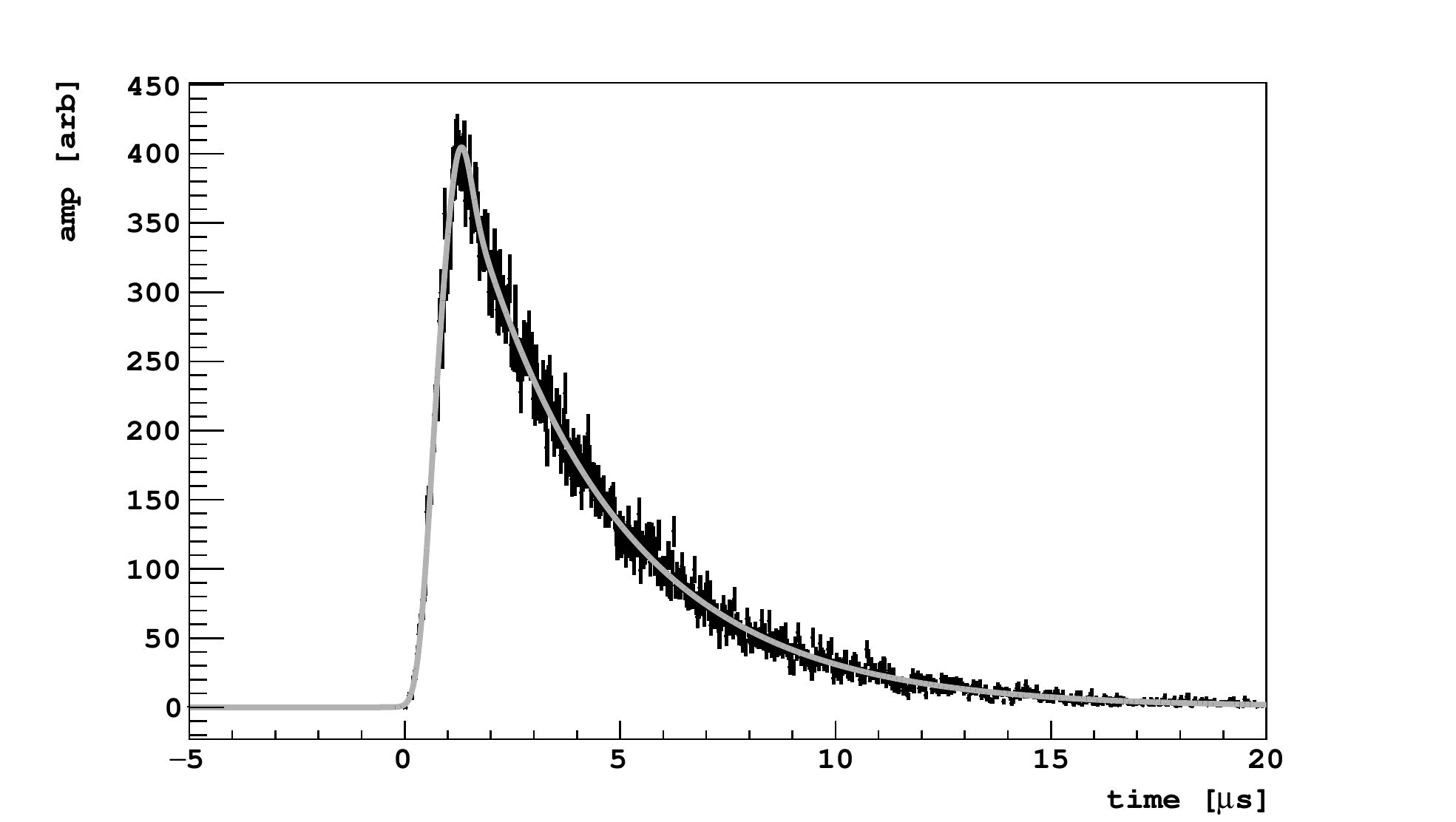}
    \includegraphics[width=\figurewidth]{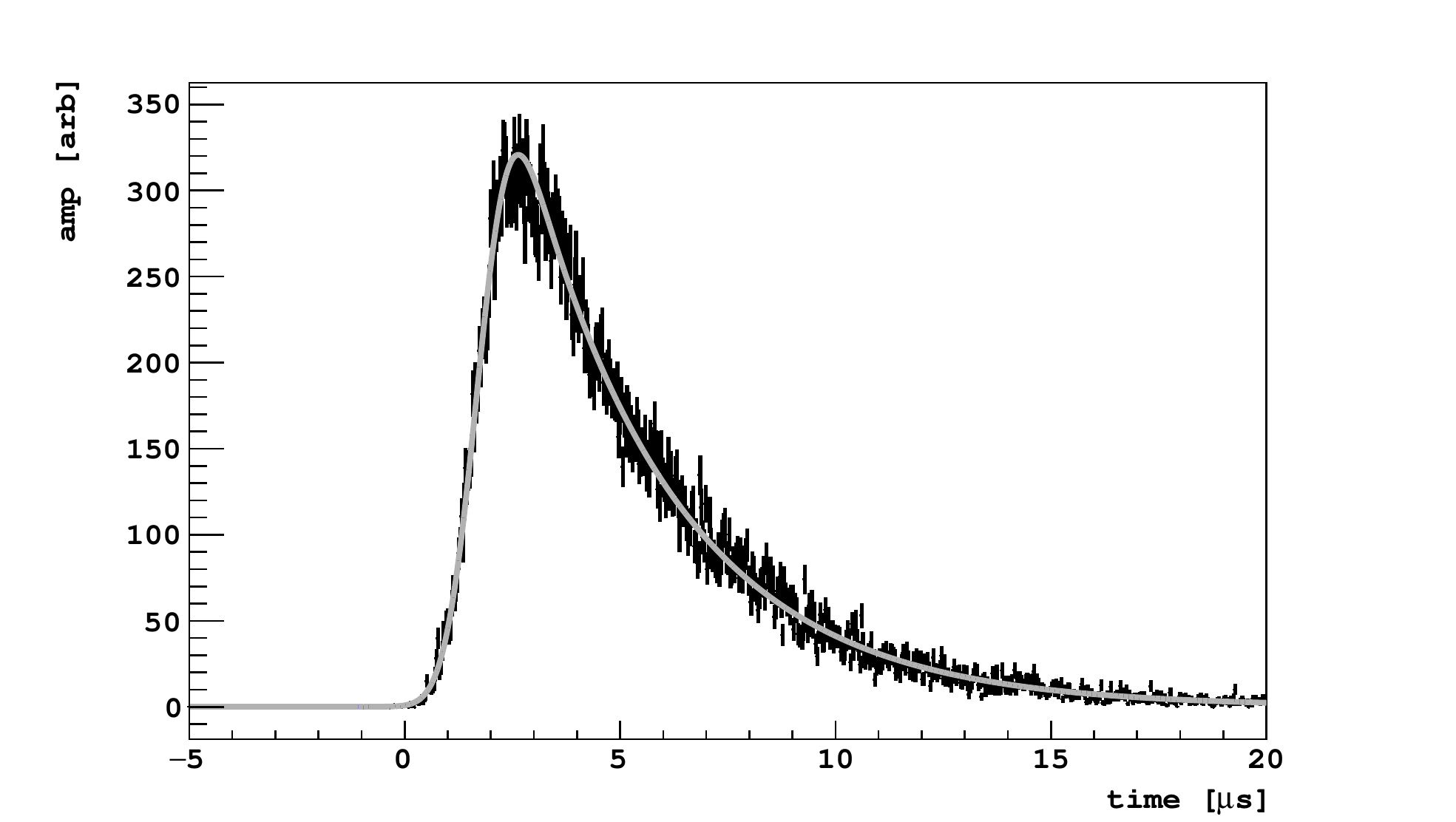}
\caption{Examples of S2 pulse shape fits for the electron diffusion measurement. Top: Event with a \SI{22}{\us} drift time. Bottom: Event with a \SI{331}{\us} drift time. The waveforms have been re-binned to \SI{32}{ns} sampling, and the x-axes redefined such that $t=0$ is at the S2 start time.}
\label{fig:diffFitExamples}
\end{figure}

\begin{figure}
\centering
\includegraphics[width=\figurewidth]{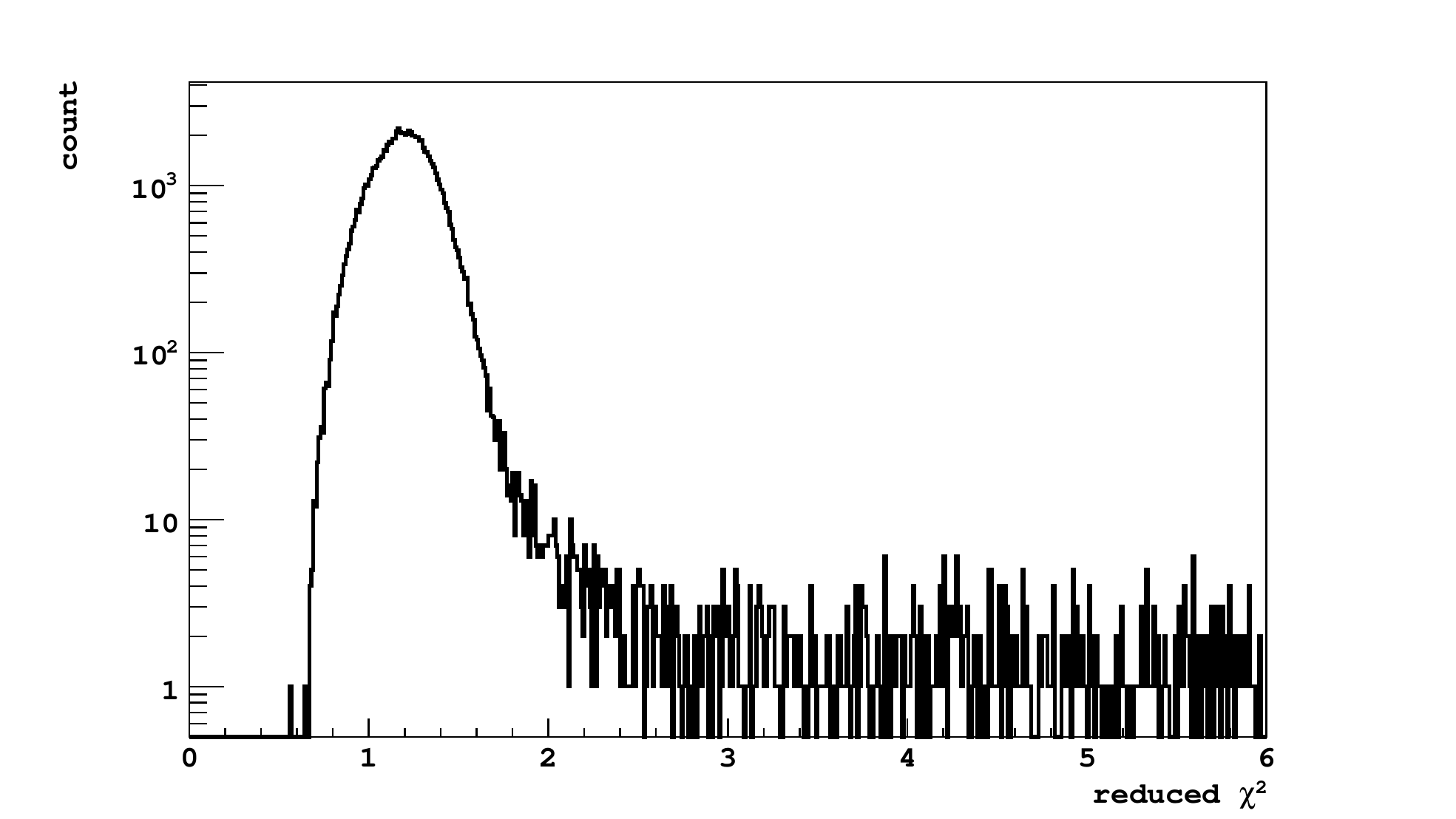}
\caption{Reduced $\chi^2$ of S2 pulse shape fits to \num{8.95e4} events in the diffusion analysis.}
\label{fig:diffChi}
\end{figure}

There are \num{8.95e4} events that pass our selection cuts. We fit the S2 pulse shape to each one. Fig.~\ref{fig:diffFitExamples} shows examples of some of the fits. 
\SI{94.5}{\percent} of the events have a reduced $\chi^2$ smaller than 1.5, as shown in Fig.~\ref{fig:diffChi}. 
To study the diffusion of the ionization electron cloud, we extract the smearing parameter $\sigma$ for each event.
First, we convert the smearing parameter from a time to a length scale, ignoring the drift-time-independent smearing ($\sigma_0$). The physical length $\sigma_L$ of the electron cloud just below the grid is related to the fit parameter $\sigma$ via Eqn.~\ref{eqn:sigma_relation}. From Eqn.~\ref{eqn:diffusion} we expect that $\sigma_L^2$ should be linear to $t_d$.
\begin{figure}
\centering
\includegraphics[width=\figurewidth]{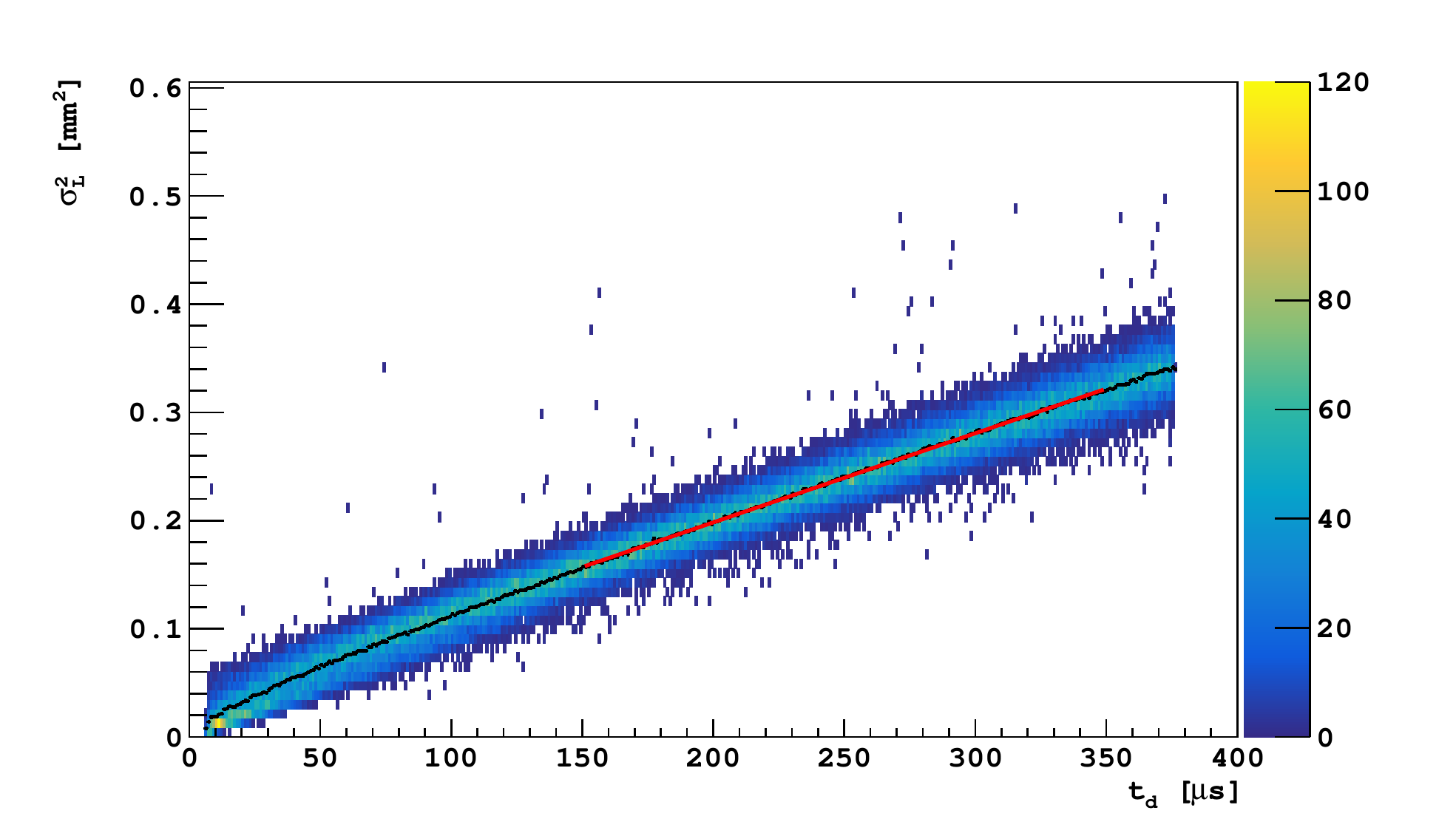}
\caption{We extract the Gaussian smearing term $\sigma$ from the S2 pulse shape fits, convert to length scale via $\sigma_L = v_d \sigma$ and plot $\sigma_L^2$ vs. drift time. The mean of the distribution is black markers and the fit of Eqn.~\ref{eqn:diff_fit} to the mean from $\SI{50}{\us}$ to $\SI{350}{\us}$ is shown as red curve. (See the web version of this article for color.)}
\label{fig:diffusion}
\end{figure}
The diffusion constant is then easily evaluated by fitting a line to the mean of the $\sigma_L^2$ vs. $t_d$ distribution:
\begin{equation} \label{eqn:diff_fit}
  \sigma_L^2 = \sigma_0^2+2D_L t_d
\end{equation}
Recall from Sec. \ref{sec:basic_shape} that the$\sigma_0$ term accounts for any systematic smearing independent of drift time, including the initial spread of the electron cloud. In DarkSide-50, $\sigma_0$ is small relative to $\sigma_L$. 

However, as evident in Fig.~\ref{fig:diffusion}, diffusion ($\sigma_L^2$) is nonlinear with respect to drift time, particularly in the region with $t_d < \SI{150}{\us}$. 
The grid mesh used in the DarkSide-50 TPC has \SI{2}{mm} pitch hexagonal cells. A COMSOL electric field simulation has shown that as electrons travel past the grid the cloud suffers a distortion that adds to the longitudinal spread of the cloud. This effect contributes to the observed nonlinearity, as smaller electron clouds suffer less distortion than larger clouds spread across multiple mesh cells. The distortion effect saturates for clouds larger than $\sigma_T=\SI{0.4}{mm}$. Performing a linear fit in the drift time range of \SIrange{150}{350}{\us} avoids the nonuniform field effect, as it restricts us to the region in which all clouds suffer the same amount of distortion. An extra $\SI{\pm 0.08}{cm^2/s}$ is assigned as systematic uncertainty to account for the nonlinearity. This uncertainty is evaluated on simulation results by changing the fit range within \SIrange{150}{350}{\us}. 

The value of the diffusion constant is sensitive to the range of $t_d$ used in the linear fit, because of the observed nonlinearity. Earlier windows tend to give a larger diffusion constant. This is also in accordance with the additional spread of the electron cloud caused by Coulomb repulsion (discussed in Sec.~\ref{sec:diffCoulomb}).  Coulomb repulsion is stronger when the electron cloud has not yet diffused, producing a larger effect in the beginning of the drift and decreasing over time.

 Using various fit windows within the $t_d$ range of \SIrange{50}{350}{\us}, we find that the diffusion constant varies by \SI{\pm 5}{\percent}. Fitting to the $t_d$ region of \SIrange{150}{350}{\us}, in which the relationship between $\sigma_0^2$ and $t_{d}$ is more approximately linear, the diffusion constant is found to be $D_L=\SI{4.12 \pm 0.09}{cm^2/s}$. The uncertainty from the fit is negligible due to the high statistics, the main contribution is from the uncertainty of the nonlinearity and electron drift velocity. The total uncertainty on $D_L$ is systematics dominated and is discussed in the following section.
We quote the results from fitting Eqn.~\ref{eqn:diff_fit} in the range of \SIrange{150}{350}{\us} without subtraction of the Coulomb repulsion effect to remain consistent with the literature. 

\subsection{Systematics}
We estimate the systematic uncertainty on the diffusion coefficient in a few different ways. As discussed, we evaluate the uncertainty arising from the nonlinear relationship between diffusion and drift time by varying the fit range applied to simulation results. We also repeat the full analysis on various data sets. We use different $r$ and S2 slices from the same set of runs used to produce the results of the previous section, as well as data taken at different extraction fields. 

\subsubsection{Vary $r$ and S2 slices}
\label{sec:diffVaryRS2}

\begin{figure}
\centering
\begin{subfigure}{\figurewidth}
  \includegraphics[width=\textwidth]{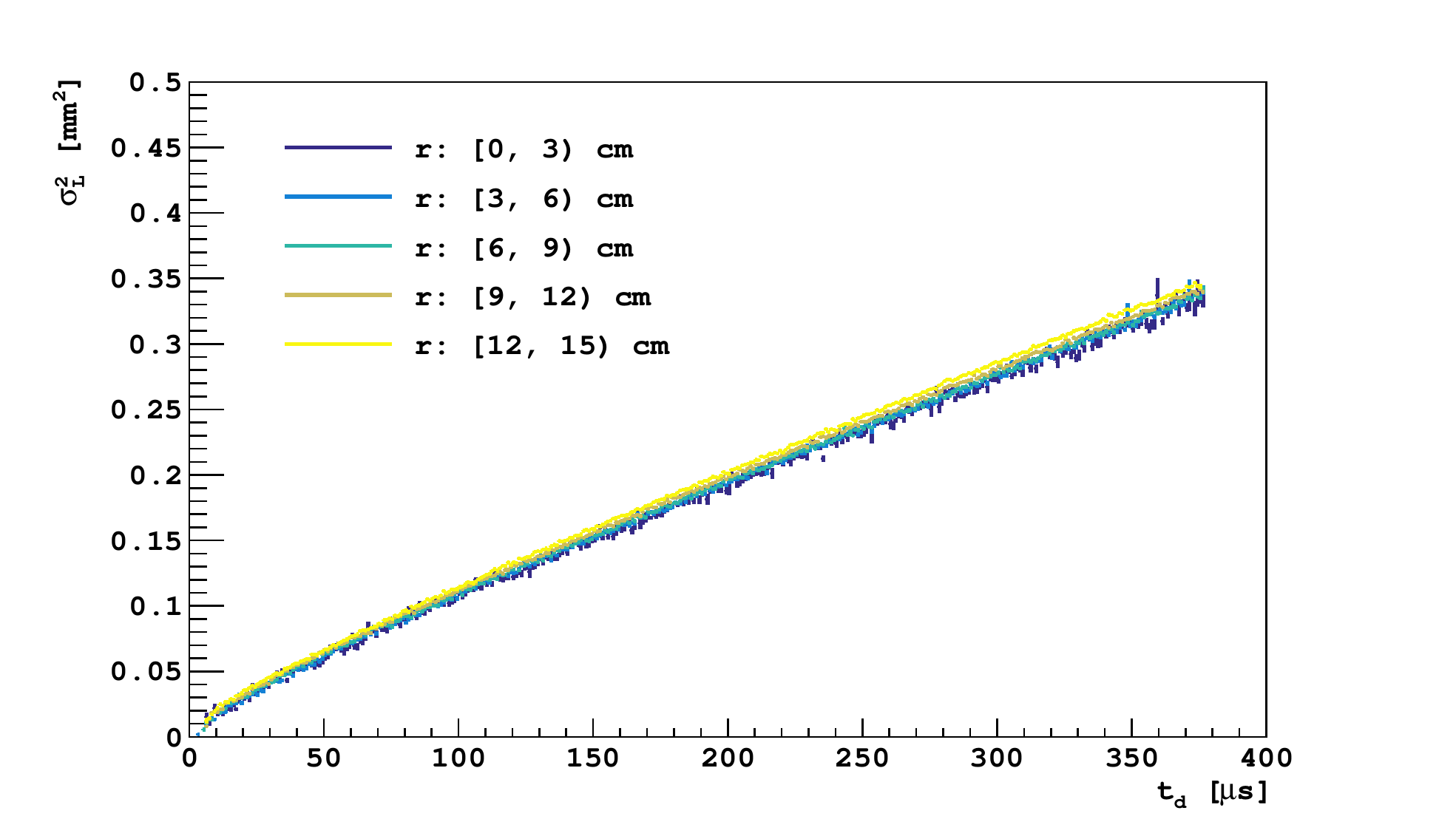}
  \caption{}
  \label{fig:diffRS2_a}
\end{subfigure}
\begin{subfigure}{\figurewidth}
  \includegraphics[width=\textwidth]{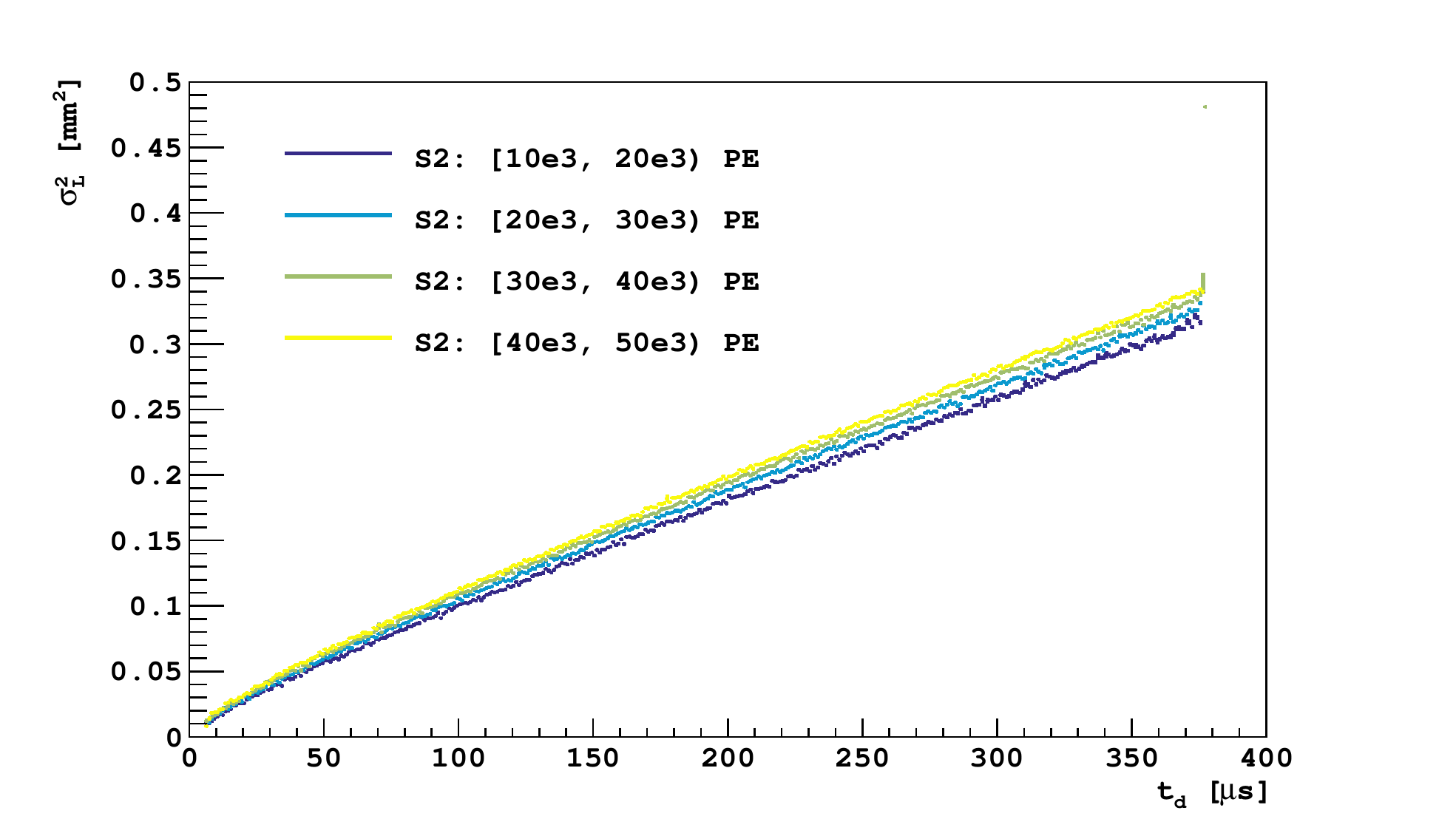}
  \caption{}
  \label{fig:diffRS2_b}
\end{subfigure}
\caption{(a) Diffusion measurement using various $r$ slices with a constant S2 slice. (b) Diffusion measurement using various S2 slices with a constant $r$ slice. (For interpretation of the references to color in this figure legend, the reader is referred to the web version of this article.)}
\label{fig:diffRS2}
\end{figure}

Ideally, $D_L$ should be independent of $r$ and S2 size. The analysis chain is applied identically to the same runs using the same cuts, but selecting events in different $r$ and S2 slices. We choose 8 additional slices:
\begin{itemize}
\item $r$ in the ranges [0,3), [3,6), [6,9), [12,15) cm all with S2 in the range [4, 5] $\times 10^4$PE.
\item S2 in the ranges [1, 2), [2, 3), [3, 4) $\times10^4$PE all with $r$ in the range [9,12) cm.
\end{itemize}
The event-by-event S2 fit procedure is identical to Sec.~\ref{sec:diffFitProcedure}, and the results are shown in Fig.~\ref{fig:diffRS2}. Only events with reduced $\chi^2\num{< 1.5}$ (\SI{94.5}{\%} of all events) are selected for all slices. The extracted diffusion constants agree to within \SI{3}{\percent} for the various $r$ slices and \SI{5}{\percent} for the various S2 slices. There is a systematic bias towards larger $D_L$ for larger $r$ and S2. 

The bias might be explained by Coulomb repulsion. Stronger repulsion drives the fitting result of $D_L$ to larger values. Events with larger S2 have a higher electron spatial density and therefore stronger self-repulsion during drift. Since the S2 light yield is lower towards the edge of the TPC \cite{Agnes:2017cz},
events with the same number of S2 photoelectrons at larger $r$ have a larger electron population than is observed, and are therefore subject to a stronger repulsion. This assumption is examined by simulation in Sec.~\ref{sec:diffCoulomb}.

\subsubsection{Vary extraction field}

\begin{figure}
  \centering
  \begin{subfigure}{0.49\textwidth}
    \includegraphics[width=\textwidth]{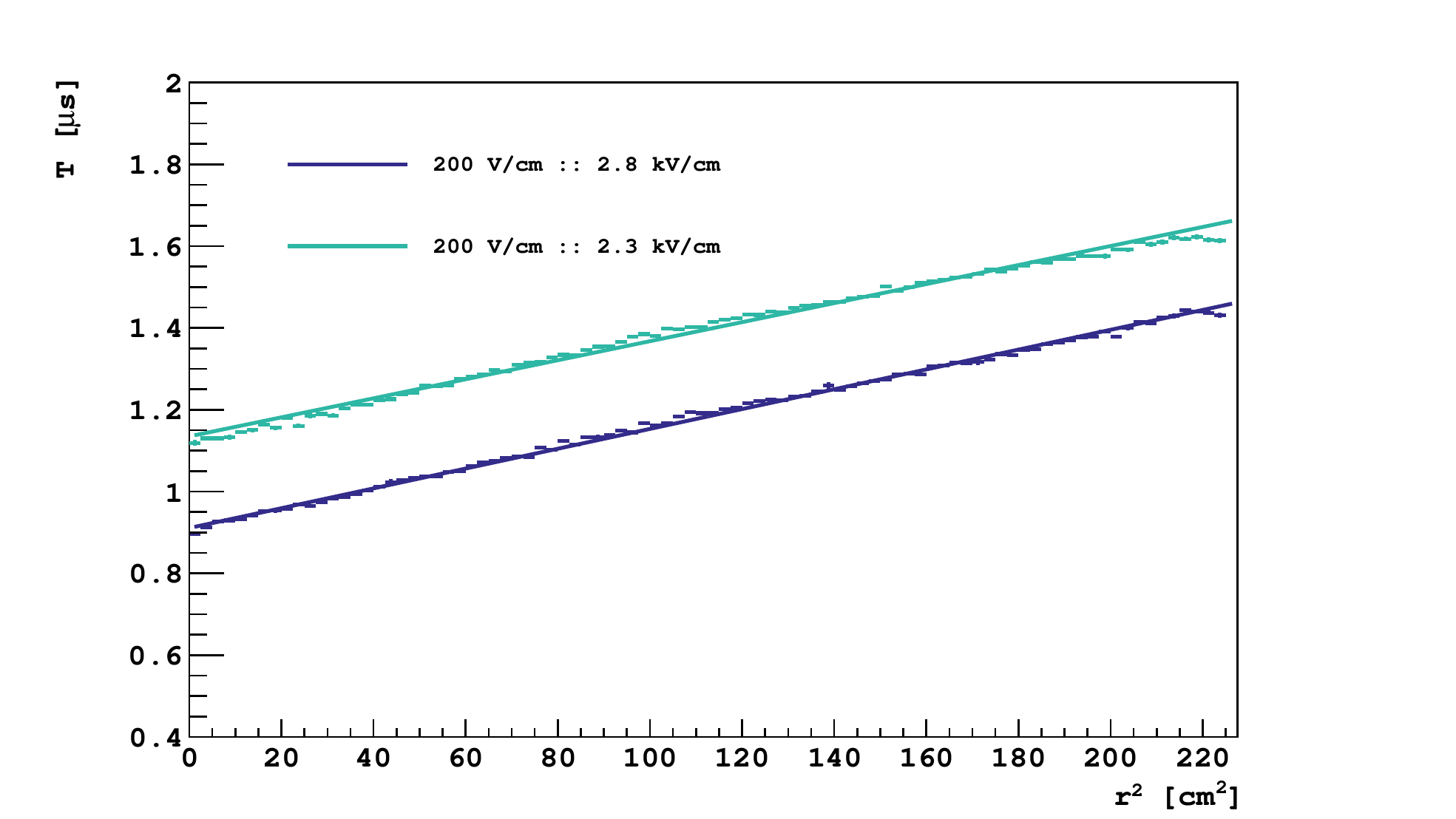}
    \caption{}
  \end{subfigure}
  \begin{subfigure}{0.49\textwidth}
    \includegraphics[width=\textwidth]{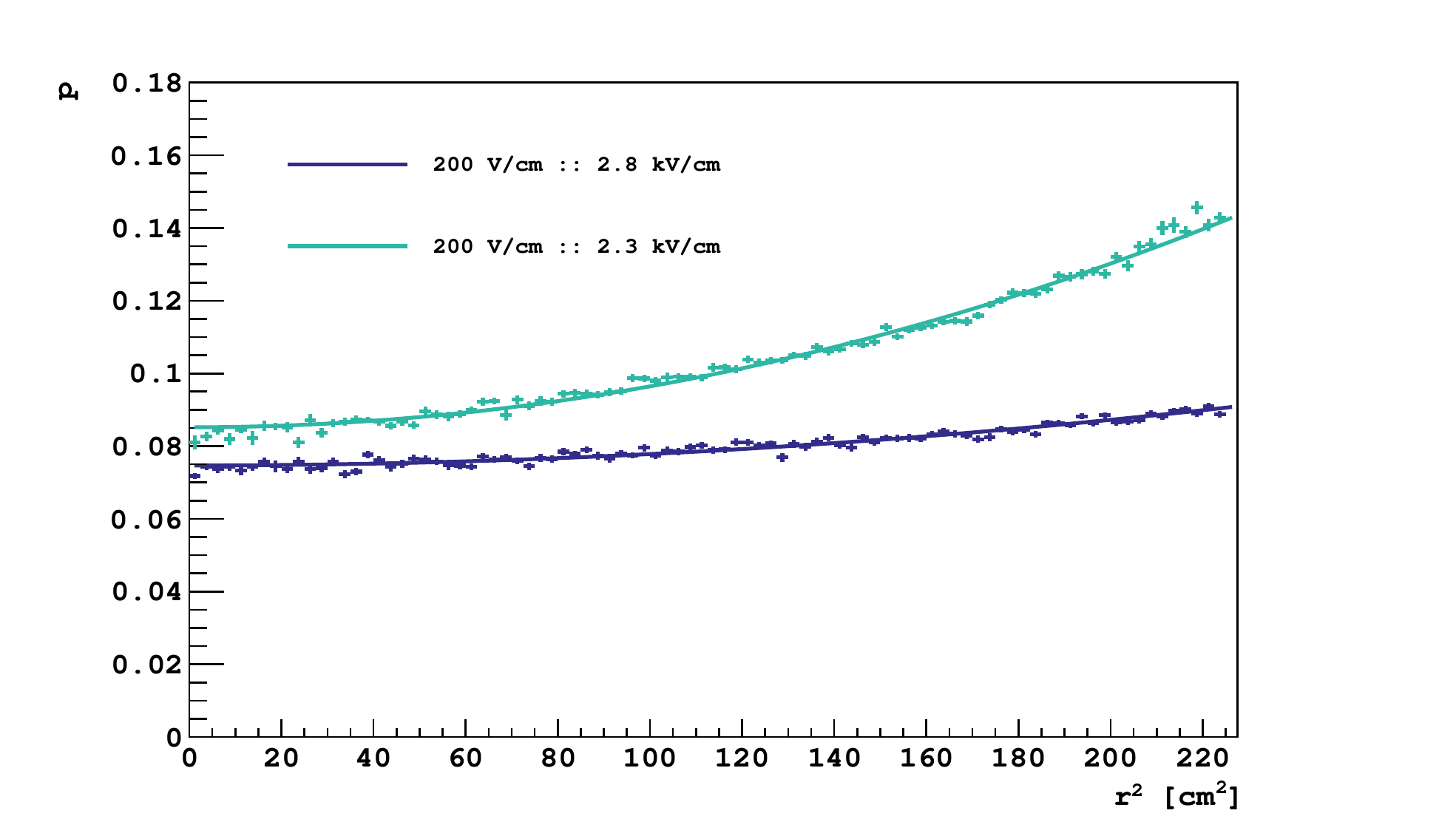}
    \caption{}
  \end{subfigure}
  \caption{Mean of (a) $T$ vs. $r^2$ and (b) $p$ vs. $r^2$ distributions for standard \SI{2.8}{kV/cm} extraction field data (\textbf{\textcolor{blue}{blue}}) and \SI{2.3}{kV/cm} extraction field data (\textbf{\textcolor{cyan}{cyan}}). (For interpretation of the references to color in this figure legend, the reader is referred to the web version of this article.)}
    \label{fig:T_R_lowExtr}
\end{figure}

Similarly, $D_L$ should be independent of the extraction field. Due to operational constraints, high statistics data were taken at only one other extraction field, \SI{2.3}{kV/cm}. We repeat the analysis chain applied to standard extraction field data, but must regenerate the $T(r)$ and $p(r)$ functions since the electron drift time across the gas pocket and the fast component fraction depend on the electroluminescence field. We repeat the analysis of Sec.~\ref{sec:parameter_degeneracy} with no modifications. The $T$ and $p$  distributions change but remain consistent with the forms of Eqn.~\ref{eqn:T_R} and~\ref{eqn:p_R}. The relevant parameters now have the values
$A_T = \SI{1.135\pm0.001}{\us}$, $B_T=\SI{488\pm 2}{cm^2}$ and
$A_p=(8.52\pm0.02)\times 10^{-2}$, $B_p=\SI{275\pm1}{cm^2}$, as shown in Fig.~\ref{fig:T_R_lowExtr}.

\begin{figure}
  \centering
  \includegraphics[width=\figurewidth]{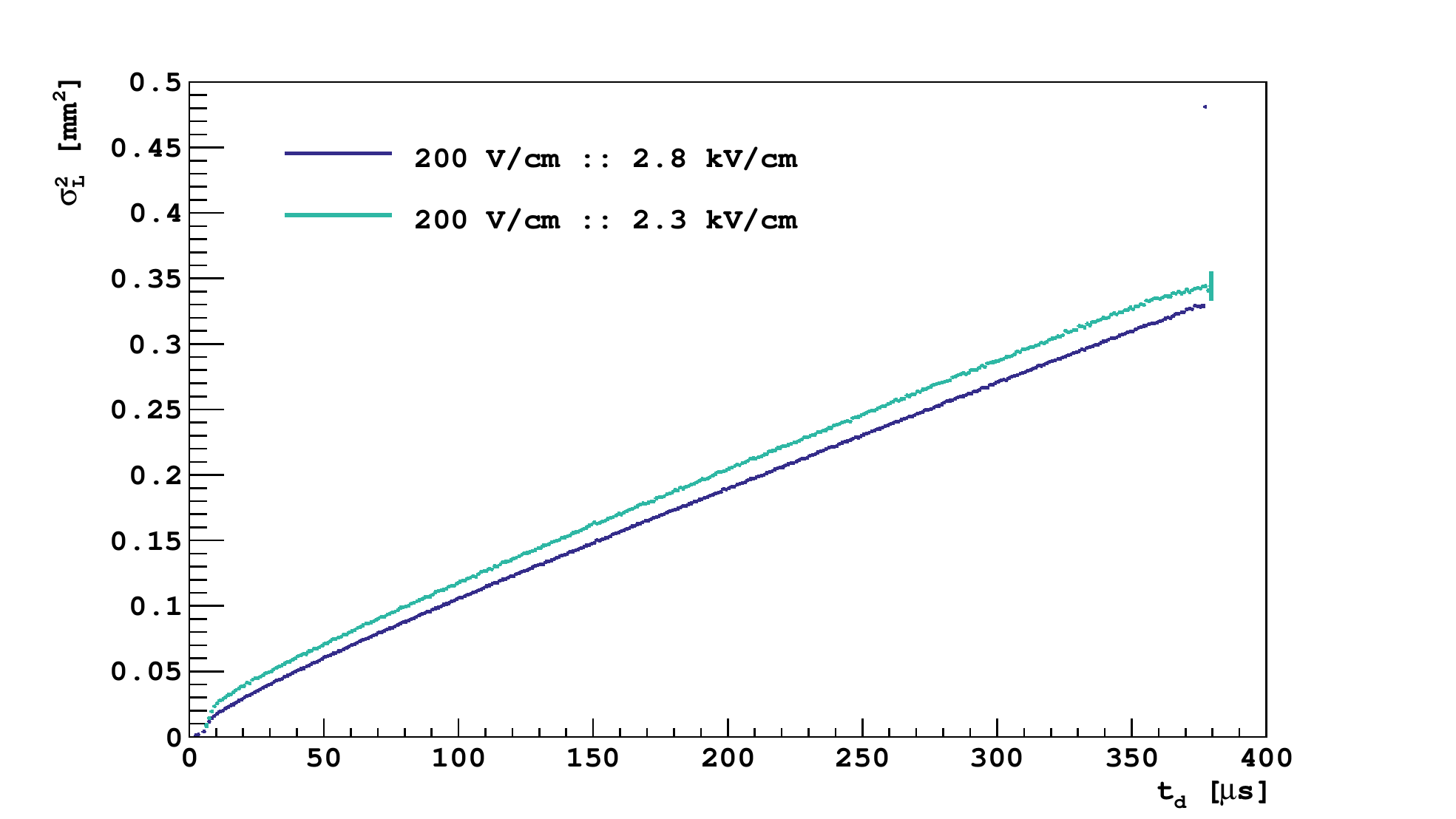}
  \caption{Mean of $\sigma_L^2$ vs. $t_d$ for  \SI{2.3}{kV/cm} extraction field data (\textbf{\textcolor{cyan}{cyan}}) and standard \SI{2.8}{kV/cm} extraction field data (\textbf{\textcolor{blue}{blue}}). (For interpretation of the references to color in this figure legend, the reader is referred to the web version of this article.)}
  \label{fig:sigma_drift_lowExtr}
\end{figure}

Using the new $T(r)$ and $p(r)$ functions, we repeat the analysis chain of Sec.~\ref{sec:diffEventSelection} and~\ref{sec:diffFitProcedure} and extract the $\sigma_L^2$ vs. $t_d$ distribution. Due to the lower statistics relative to standard field data, we extend the $r$ and S2 slices to include \SIrange {0}{18}{cm} and $(1-5)\times\SI{e4}{PE}$, respectively. With the reduced electroluminescence field, we are probing a higher range of event energies. The mean of the resulting $\sigma^2$ vs. $t_d$ distribution is shown in Fig.~\ref{fig:sigma_drift_lowExtr}.
We see that there is an overall shift in the distribution, which is expected since, with the lower electroluminescence field, the electrons are more slowly extracted from the LAr surface and drifted in the gas. The slope, and therefore also $D_L$, is consistent with the results of other data sets. 

\subsubsection{Summary of systematics}

The values of the longitudinal diffusion constant extracted from the various data sets are summarized in Tab.~\ref{tab:systematics}. 
\begin{table*}
  \centering
  \begin{tabular}{ c  c  c  c  c  c }
    Drift [\si{V/cm}] & Extr. [\si{kV/cm}] & R [\si{cm}] & S2 [\SI{e3}{PE}] & $D_L$ [\si{cm^2/s}] & $\sigma_0^2$ [$\times\SI{e-2}{mm^2}$] \\
    \hline
    200 & 2.8 & [0, 3] & [40, 50]     & $4.09\pm0.05$ & $2.94\pm0.10$ \\
    200 & 2.8 & [3, 6] & [40, 50]     & $4.10\pm0.04$ & $2.98\pm0.07$ \\
    200 & 2.8 & [6, 9] & [40, 50]     & $4.10\pm0.04$ & $3.07\pm0.06$\\
    200 & 2.8 & [9, 12] & [40, 50]   & $4.12\pm0.04$ & $3.34\pm0.06$\\
    200 & 2.8 & [12, 15] & [40, 50] & $4.19\pm0.04$ & $3.45\pm0.06$ \\
    \hline
    200 & 2.8 & [9, 12] & [30, 40] & $4.09\pm0.04$ & $3.00\pm0.05$ \\
    200 & 2.8 & [9, 12] & [20, 30] & $4.00\pm0.04$ & $2.81\pm0.05$ \\
    200 & 2.8 & [9, 12] & [10, 20] & $3.92\pm0.04$ & $2.37\pm0.05$ \\
    \hline
    200 & 2.3 & [0, 15] & [10, 50] & $4.16\pm0.04$ & $3.76\pm0.07$ \\
  \end{tabular}
  \caption{ A summary of the diffusion constant values $D_L$ measured from different data sets and different extraction fields. Errors reflect the fitting uncertainty and uncertainty from drift velocity in Tab. \ref{tab:vdrift}}
  \label{tab:systematics}
\end{table*}
The given uncertainties on $D_L$ are dominated by the uncertainty in the drift velocity. The uncertainty on $\sigma_0$ is attributable to statistical uncertainties and the systematics introduced by fixing the fitting parameters $T(r)$ and $p(r)$. We obtain an average value of the diffusion constant by weighting the measured $D_L$ from different $r$ slices with the number of events in each slice, giving equal weight per unit S2 energy, and finally giving equal weight to the two extraction fields. The result is $D_L=\text{\SI{4.09(12)}{cm^2/s}}$, where the uncertainty is dominated by systematics arising from variations in S2 size, radius ($R$), extraction field, and the uncertainty from nonlinearity.

\subsection{Coulomb repulsion}
\label{sec:diffCoulomb}
In Tab.~\ref{tab:systematics} and Fig.~\ref{fig:diffusion}, we observe that the longitudinal diffusion constant is systematically growing with S2 and $\sigma^2_L$ is not strictly linear with $t_d$ as expected from Eqn.~\ref{eqn:diff_fit}. These observations can be at least partially explained by the effect of Coulomb repulsion between the electrons during drift. Adopting a similar approach as~\cite{Shibamura:1979fb}, we simulate the distribution of electrons undergoing both diffusion and Coulomb repulsion to examine this effect.

%simulation method
After the primary ionization and recombination process, we assume that the electron cloud that separated from positive ions has a Gaussian spatial distribution with an appropriate initial spread (\SI{30}{\um}), which is estimated based on simulation results from G4DS \cite{Agnes:2017cz}. During drift, the electric field at each electron is dominated by the drifting field, so the repulsive movement of an electron relative to the center of the electron cloud is
\begin{equation}
	\mathbf{v}_r = (\mathbf{E}-\mathbf{E}_d)\mu = \mathbf{E}_r\mu
\end{equation}
where $\mathbf{E}_d$ is the drift field and $\mathbf{E}_r$ is the repulsive field generated by the other electrons in the cloud according to Coulomb's law, and $\mu$ is the electron mobility, which is assumed to be constant as $E_r \ll E_d$. 
In each \SI{0.5}{\us} time interval, ignoring the difference between $D_L$ and $D_T$ in Eqn.~\ref{eqn:diff_full}, electrons take a random walk according to the diffusion constant $D_L$ and a repulse given by $\mathbf{E}_r$ at that point. 
\begin{equation}
	\Delta\mathbf{r} = \Delta t \mathbf{v}_r + \Delta \mathbf{r}_d
\end{equation}
where $\Delta \mathbf{r}_d$ is a random vector following a 3D Gaussian distribution with isotropic variance $\sigma^2 = 2D_L\Delta t$. That is to say, for simplicity we assume the diffusion is isotropic. 
The distribution of the electron cloud will be distorted slightly away from a Gaussian by the Coulomb force, so we use the RMS of electron positions along the z direction in place of the standard deviation, $\sigma_L$, in Eqn.~\ref{eqn:diff_fit}.
As the electron number in each cloud is on the order of \SI{e3}, random fluctuations are large after many time intervals. The final result is averaged over an ensemble of $2\times10^5$ simulated events. 

\begin{figure}
 \centering
  \begin{subfigure}{\figurewidth}
    \includegraphics[width=\textwidth]{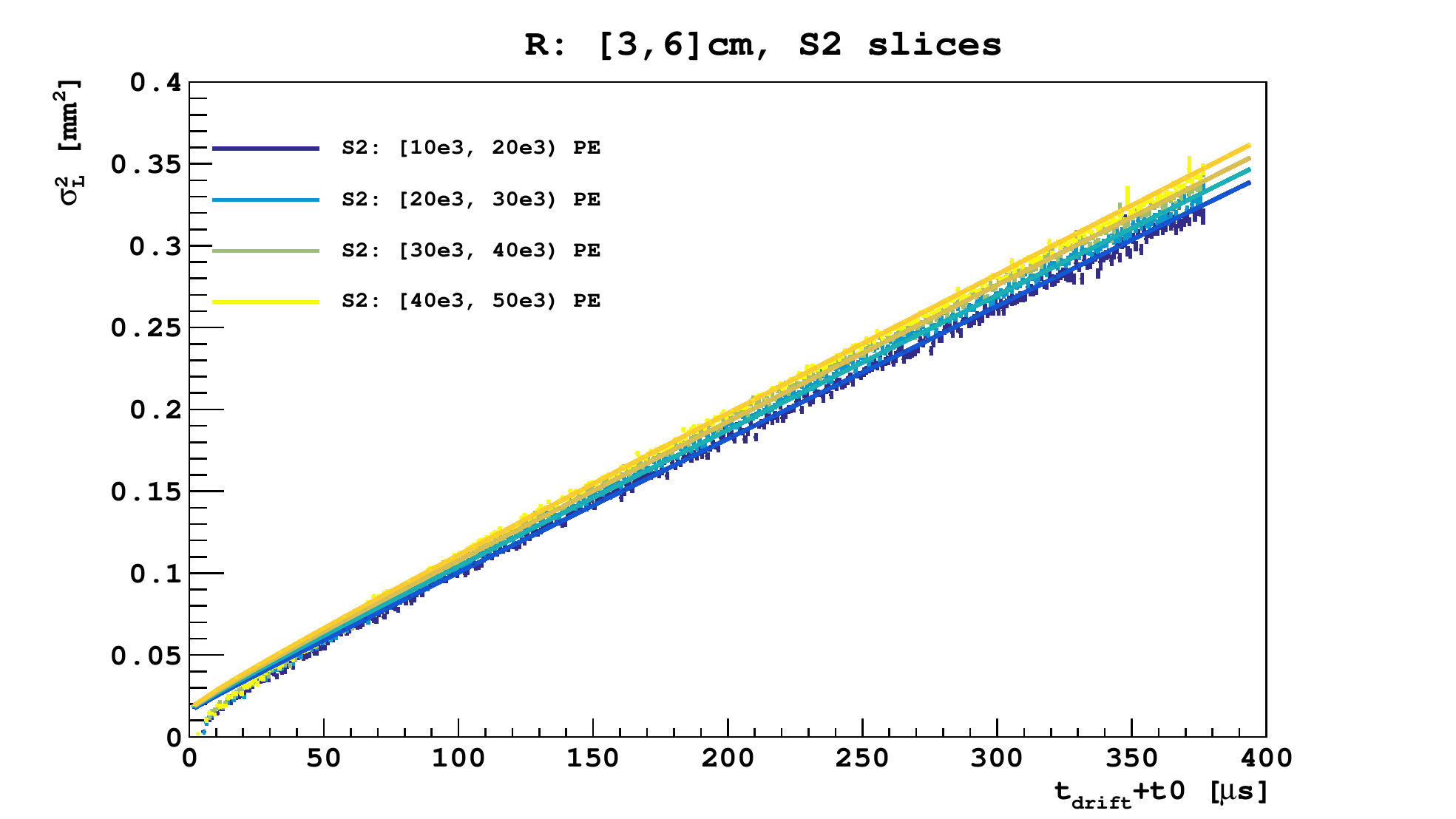}
    \caption{}
  \end{subfigure}
  \begin{subfigure}{\figurewidth}
    \includegraphics[width=\textwidth]{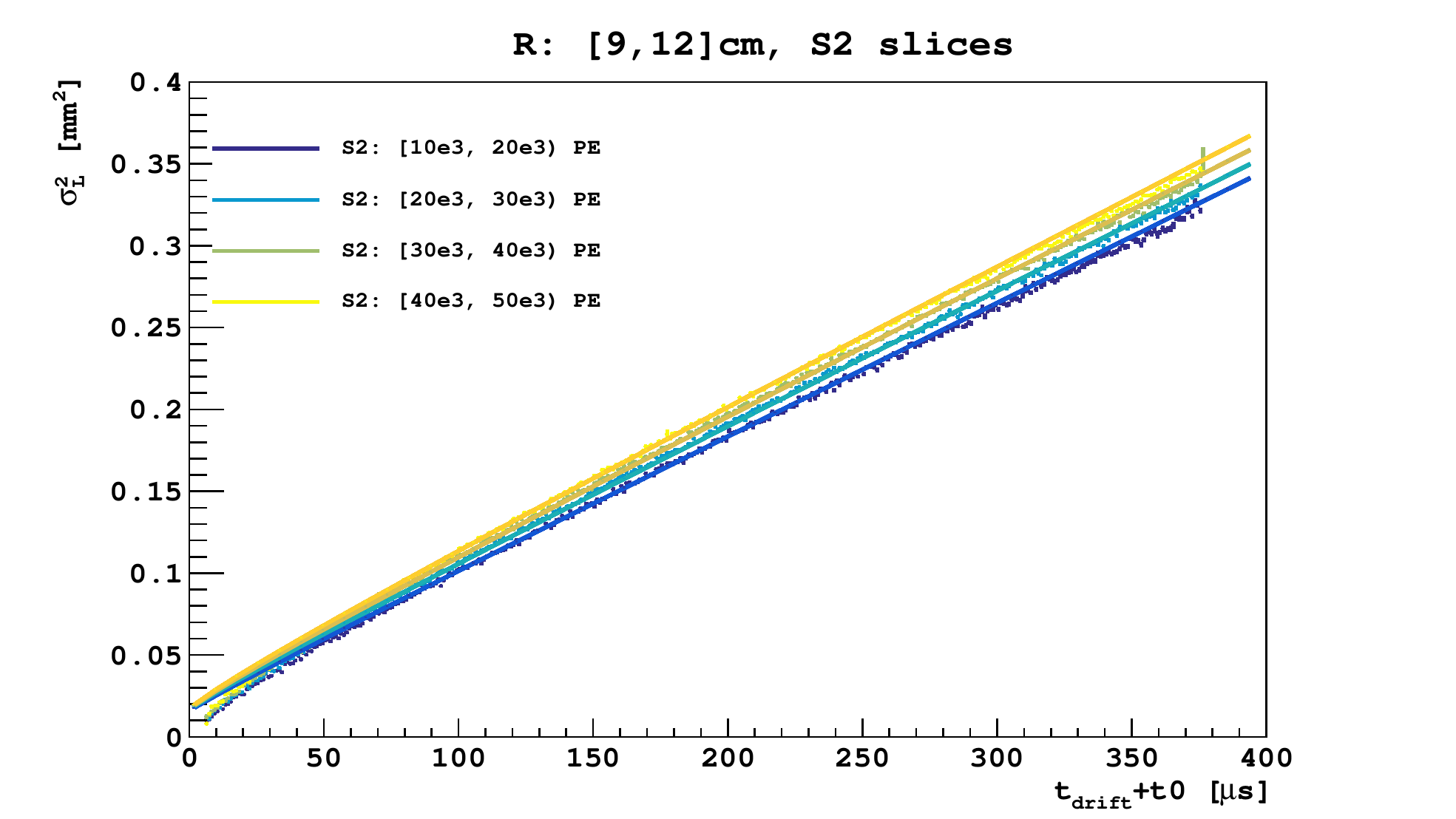}
    \caption{}
  \end{subfigure}
  \caption{Simulation results of electron diffusion with self Coulomb repulsion (lines) compared to data from Fig.~\ref{fig:diffRS2_b} (points): (a) Events with $r$ = [3, 6] cm, $D_L = \SI{3.88}{cm^2\per\second}$, $\sigma_0 = \SI{1.20e-4}{cm^2}$, $Y_{S2} = \SI{13.5}{PE\per e}$. (b) Events with $r$ = [9, 12] cm, $D_L = \SI{3.88}{cm^2\per\second}$, $\sigma_0 = \SI{1.20e-4}{cm^2}$, $Y_{S2} = \SI{11.5}{PE\per e}$. Systematic dependence of $D_L$ and $\sigma_0$ on $r$ can be decoupled by introducing Coulomb repulsion and a $r$-dependent S2 yield. (For interpretation of the references to color in this figure legend, the reader is referred to the web version of this article.)}
  \label{fig:diffCoulomb}
\end{figure}

Since the TPC does not measure charge directly, we take the S2 PE yield per drifting electron as a tuning parameter while assuming that the yield is constant within the energy range $(1-5)\times\SI{e4}{PE}$. Finally, we tune the simulation to the 4 data distributions shown in Fig.~\ref{fig:diffRS2_b} using 3 parameters: the longitudinal diffusion constant $D_L$, the S2 PE yield ($Y_{S2}$, defined as the detected number of PE per electron drifted to gas pocket), and a constant to account for any other systematic drift time-independent smearing ($\sigma_0$). The results are shown in Fig.~\ref{fig:diffCoulomb}.

Diffusion curves at different S2 energy and $r$ slices can be fit well with the same $D_L$ and $\sigma_0$ while only tuning $Y_{S2}$. After decoupling the systematic influence of radius on S2 yield, we get $D_L = \SI{3.88 \pm 0.05 }{cm^2\per\second}$. The uncertainty comes from the statistics of the simulation results. 
This number is systematically smaller than the results in Tab.~\ref{tab:varyDrift}, which is to be expected as Coulomb repulsion contributes to the spread of electrons in a drifting electron cloud. The paper published by the ICARUS collaboration also pointed out this bias~\cite{Cennini:1994ba}. $Y_{S2}$ decreases with increasing radius in the simulation results, in agreement with the other studies of S2 yield in DarkSide-50 \cite{Agnes:2017cz}. 
Unfortunately, in order to match the energy dependence observed in the data we require an S2 yield that is $\sim$2 times lower than has been measured through independent calibration analyses (not published). Restricting our S2 yield to the measured value cannot replicate the S2-dependence that we see in the data. The simulation was also replicated with initial electron distributions exhibiting some spread in either the longitudinal or transverse direction, but the results were not sufficient to resolve the discrepancy in $Y_{S2}$. Due to this discrepancy, we do not include the Coulomb repulsion effect when reporting our final result.

\subsection{Comparison to literature}
\label{sec:literatureComparison}

In order to make a reasonable comparison of the measured longitudinal diffusion constant to literature, we define the effective electron energy, $\epsilon_L$ \cite{Atrazhev:1998hl}. At low drift electric fields as in this study, the electrons are thermal (i.e. have nearly no extra energy from the field. Previous studies have shown that electrons start heating above \SI{200}{V/cm} in LAr \cite{Atrazhev:1998hl, Huang:1981fg,Gushchin:1982wu}). It is interesting to note that the relationship between electron temperature and electric field strength in liquid xenon is much stronger. As seen in Ref. \cite{Atrazhev:2005ab}, the electron temperature rises dramatically with field, even at field strengths lower than considered here ($<$ 100 V/cm).

At the drift fields considered in this analysis for liquid argon, diffusion of the electron cloud should follow the Einstein-Smoluchowski diffusion equation
\begin{equation}
D_L = \frac{kT}{e} \mu
\label{einstein}
\end{equation}
where $k$ is the Boltzmann constant, $T$ is the temperature of the medium, and $e$ is the charge of the electron. 
In higher drift fields, drifting electrons are no longer thermal. 
The effective electron energy associated with longitudinal diffusion can then be defined as
\begin{equation}
\epsilon_L = \frac{D_L}{\mu}
\end{equation}
At low drift field $\epsilon_L$ should be approximately $kT/e$. In this study $T=\SI{89.2}{K}$ and $kT = \SI{7.68}{meV}$ 

\begin{figure}
\centering
\begin{subfigure}{\figurewidth}
  \includegraphics[width=\textwidth]{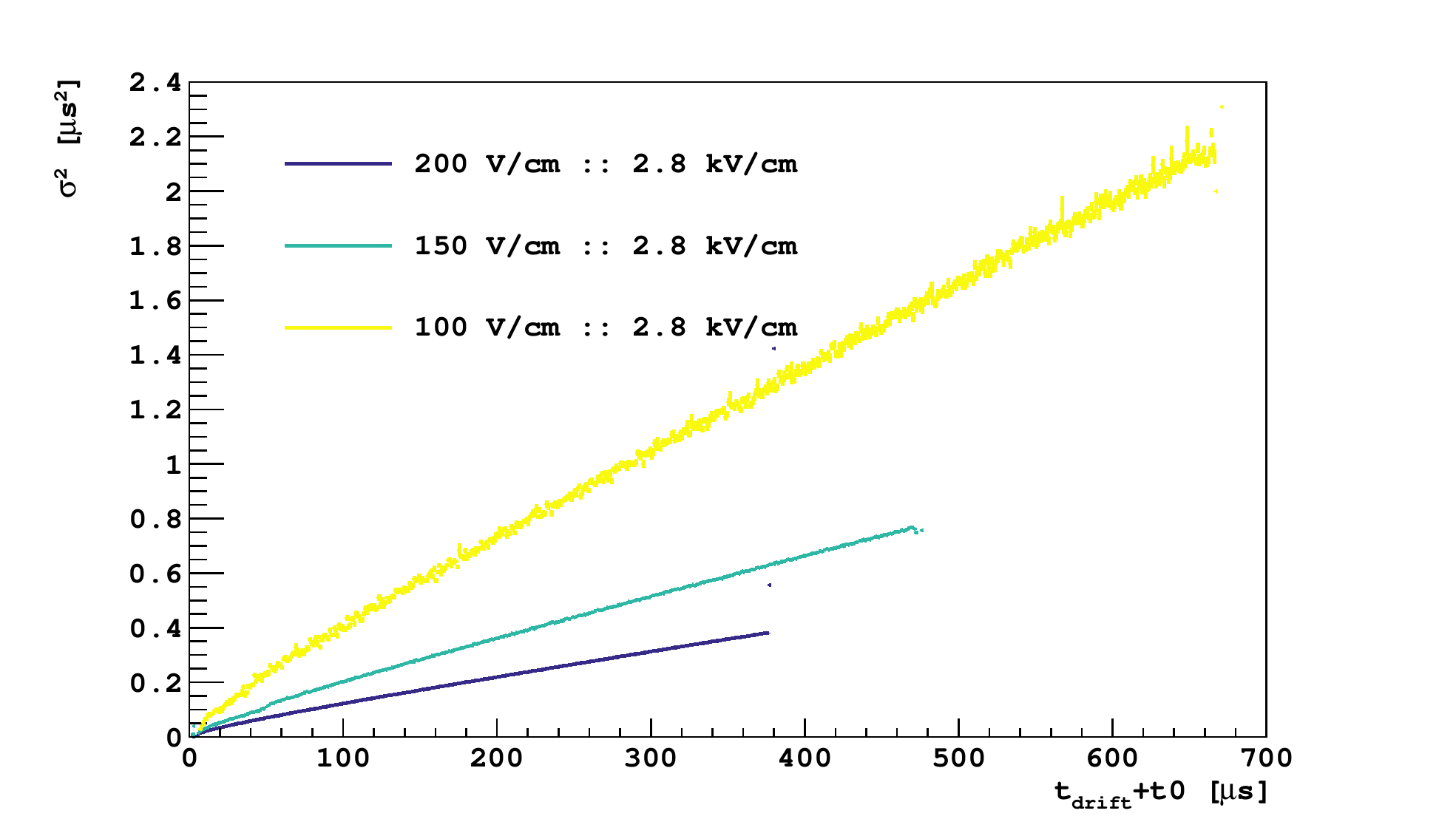}
  \caption{}
\end{subfigure}
\begin{subfigure}{\figurewidth}
  \includegraphics[width=\textwidth]{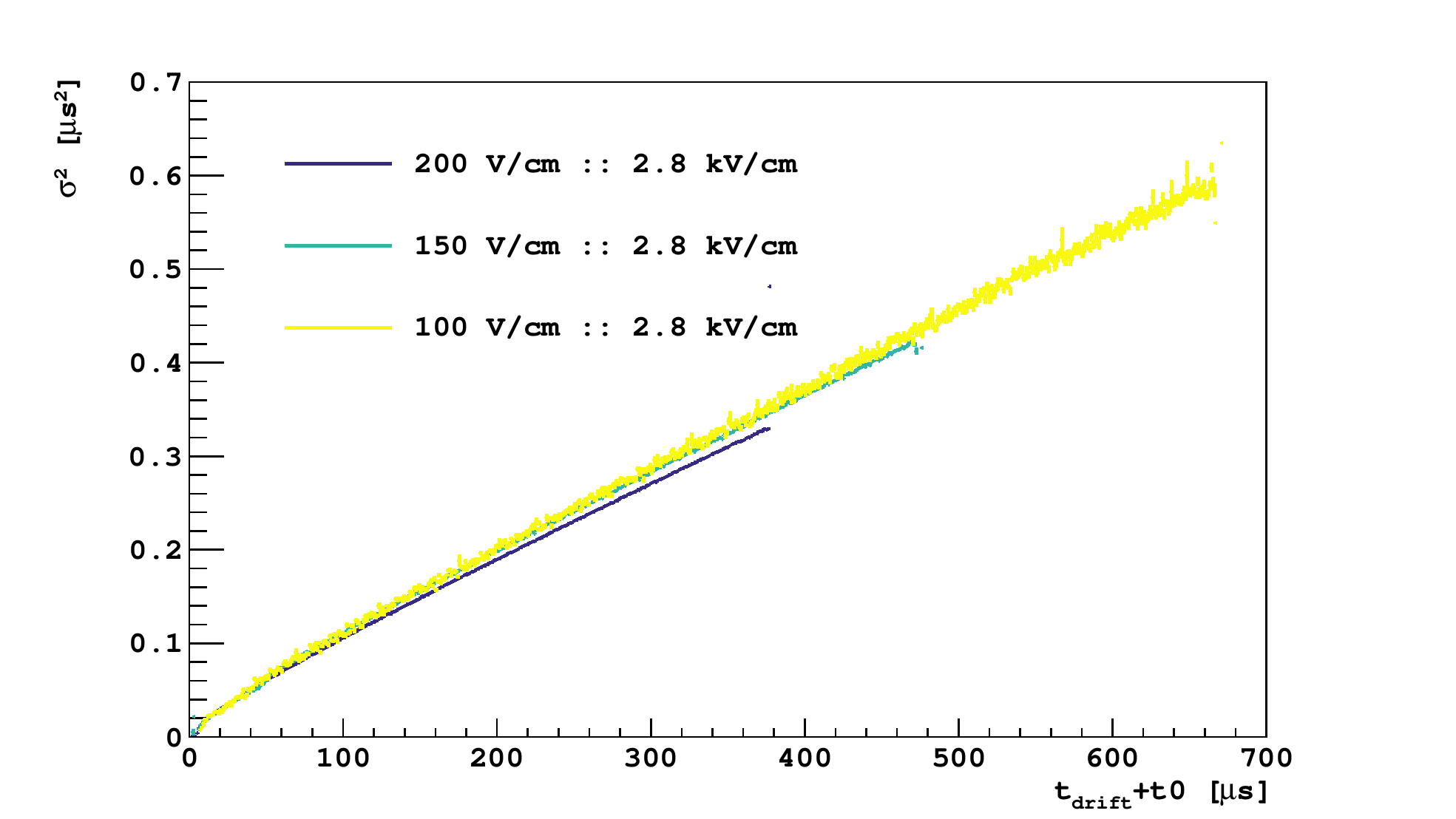}
  \caption{}
\end{subfigure}
\caption{(a) Results of the diffusion measurement for data at different drift fields. (b) After normalizing for drift velocity. (For interpretation of the references to color in this figure legend, the reader is referred to the web version of this article.)}
\label{fig:varyDrift}
\end{figure}

\begin{figure}
  \centering
  \includegraphics[width=\figurewidth]{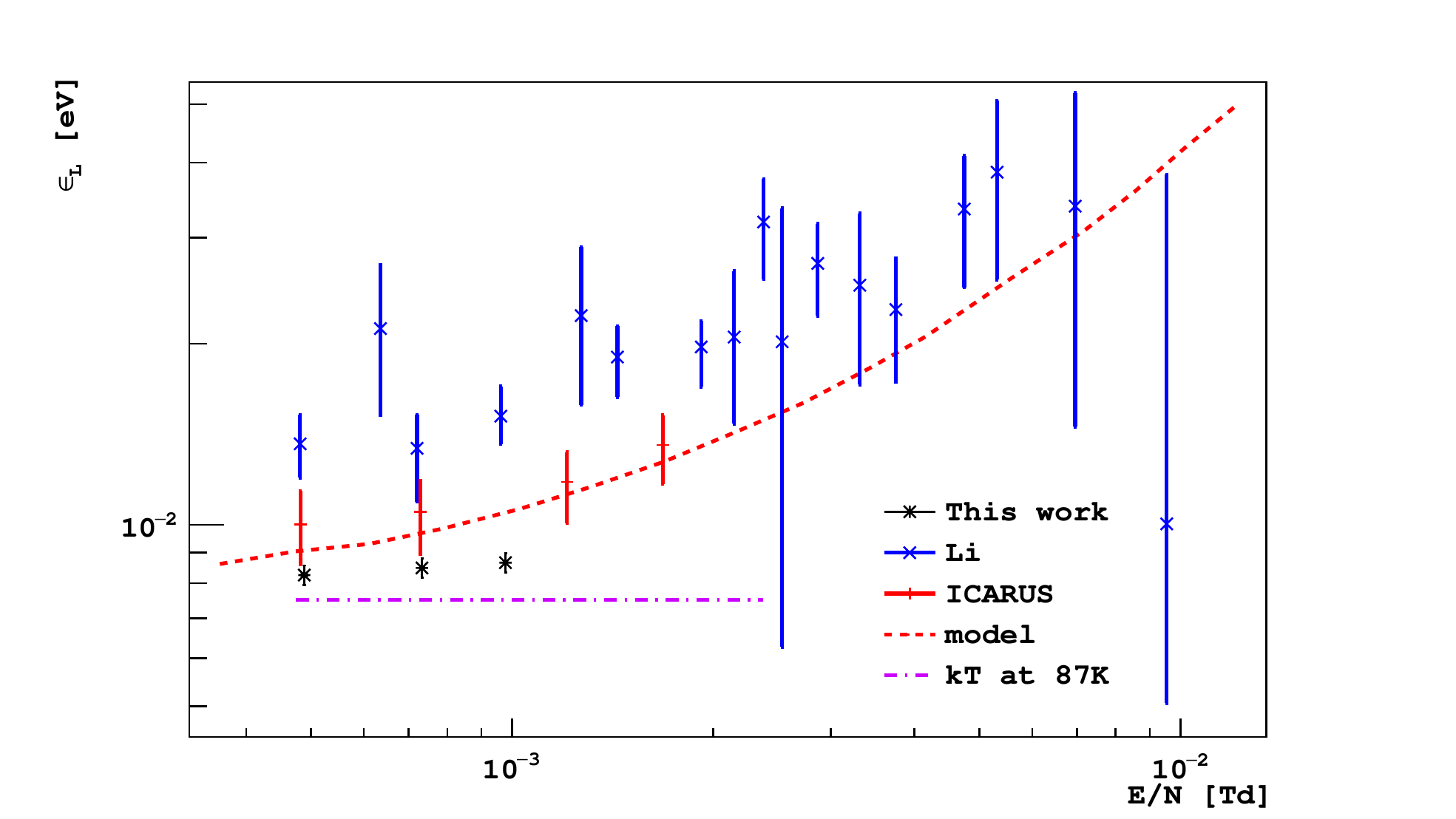}
  \caption{Electron characteristic longitudinal energy $\epsilon_L$ vs. reduced field (Td = 10$^{-17}$ V cm$^2$). Li data is from~\cite{Li:2016dz} and ICARUS data are extracted from~\cite{Cennini:1994ba}. The model is that of Atrazhev and Timoshkin~\cite{Atrazhev:1998hl}. The horizontal dashed line represents the thermal energy at \SI{87}{K}. Error bars are mainly attributable to systematics, including the uncertainty from the nonlinear relation, which is not included in the errors in the other works.}
  \label{fig:eps_L}
\end{figure}

We repeat our analysis on atmospheric argon background data taken at two different drift fields, \SI{100}{V/cm} and \SI{150}{V/cm}, to compare to the nominal \SI{200}{V/cm} drift field data. All data are taken with \SI{2.8}{kV/cm} extraction field. The event selection criteria are nearly identical to those used in the main analysis. However, due to reduced statistics in lower drift field data set, we take a wider slice in the $r$ vs. S2 plane: for all 3 drift fields, we use $r$ in the range \SIrange{0}{15}{cm}, and S2 in the range $(1-5)\times \SI{e4}{PE}$. The event-by-event fit procedure is identical to that of the standard drift field data. In particular, since the electroluminescence field is unchanged, we use the same $T(r)$ and $p(r)$ functions given by Eqn.~\ref{eqn:T_R} and~\ref{eqn:p_R}, respectively. The results are shown in Fig.~\ref{fig:varyDrift}. 
Results of the linear fit of Eqn.~\ref{eqn:diff_fit} to the points in Fig.~\ref{fig:varyDrift} are shown in Tab.~\ref{tab:varyDrift}. Error estimation is the same as for the previous analysis. Besides the uncertainties in the table, we assign the same total systematic error to the values, which are shown in Fig \ref{fig:eps_L}
\begin{table*}
  \centering
  \begin{tabular}{ c  c  c  c  c  c }
    Drift [\si{V/cm}] & Extr. [\si{kV/cm}] & R [\si{cm}] & S2 [\SI{e3}{PE}] & $D_L$ [\si{cm^2/s}] & $\sigma_0^2$ [$\times\SI{e-2}{mm^2}$] \\
    \hline
    100 & 2.8 & [0, 15] & [10, 50] & $4.35\pm0.05$ & $2.67\pm0.09$ \\
    150 & 2.8 & [0, 15] & [10, 50] & $4.21\pm0.04$ & $2.99\pm0.05$ \\
    200 & 2.8 & [0, 15] & [10, 50] & $4.05\pm0.04$ & $2.76\pm0.04$ \\
  \end{tabular}
  \caption{Diffusion constant $D_L$ measured under different drift fields. Only the uncertainty from the fit results and the drift velocity are reported.}
  \label{tab:varyDrift}
\end{table*}

We evaluate $\epsilon_L$ separately for each drift field using the appropriate mobility value from Tab.~\ref{tab:vdrift} and the measured $D_L$ without subtraction of the Coulomb repulsion effect. Results are shown in Fig.~\ref{fig:eps_L}, along with results from other experiments and models. All data points representing experimental measurements are normalized to \SI{87}{K} assuming a linear $T$ dependence of $\epsilon_L$ at very low drift field. 
The curve represents the model of Atrazhev and Timoshkin \cite{Atrazhev:1998hl}, which is calculated based on a variable phase method near the argon triple point (\SI{83.8}{K}). The data from Li et al \cite{Li:2016dz} was taken using electrons generated from an Au photocathode excited by a picosecond laser with a beam size of \SI{1}{mm} at \SI{87}{K}, while the ICARUS \cite{Cennini:1994ba} data was taken with cosmic muon tracks with a minimum ionizing particle density of $(4-5.5)\times\SI{e3}{e/mm}$ at \SI{92}{K}. The uncertainty of ICARUS data is calculated by the same method as described by Li et al. The electron density reported by Li et al is even lower than ICARUS. Neither work implements a correction based on the Coulomb repulsion effect.

The results from literature are systematically higher than the results from this work, but our measurement is closer to the thermal energy. We should note here that the data in~\cite{Li:2016dz} were taken over drift lengths between \SI{2}{cm} and \SI{6}{cm}, which corresponds to the $0-60\si{\us}$ region in our Fig.~\ref{fig:diffRS2} or~\ref{fig:sigma_drift_lowExtr} where the non-linearity is most significant. As both setups consist of a field cage with shaping rings and a grid electrode to apply an extraction field (named collection field in~\cite{Li:2016dz}), it is reasonable to expect a higher diffusion constant from a linear fit to the short drift time region in Li's study. The discrepancy between our results and the thermal energy might come from electron heating caused by the drift field. The increase in $\epsilon_L$ with drift field is discernible with the given uncertainty, indicating that the drifting electrons in the $\SI{100}{V/cm}$ to $\SI{200}{V/cm}$ drift field range is not completely thermal.

\section{Summary}

We have performed a precise measurement of the longitudinal electron diffusion constant in liquid argon using the DarkSide-50 dual-phase TPC. Radial variation of the electroluminescence field induces a strong radial dependence in the S2 pulse shape, particularly the time to the peak of the pulse, $T$, and the fast component fraction, $p$. This radial variation is accounted for by determining $T(r)$ and $p(r)$ using events from the uppermost layer of the liquid where diffusion is negligible.

The measured longitudinal diffusion constant is $\SI{4.12\pm0.09}{cm^2/s}$ for a selection of $\SI{140}{keV}$ electron recoil events subject to a $\SI{200}{V/cm}$ drift field at $\SI{89.2}{K}$. To study the systematics of our measurement we examined datasets of varying event energy, field strength, and detector volume yielding a weighted average value for the diffusion constant of \SI{4.09 \pm 0.12 }{cm^2/s}, where the uncertainty is systematics dominated. Results at all examined drift fields are systematically lower than other measured values in literature, but closer to the prediction of the Einstein-Smoluchowski diffusion equation, assuming thermalized electrons. Coulomb repulsion within the drifting electron cloud might contribute to a larger diffusion constant. However, from simulation results we conclude that the Coulomb repulsion effect might not fully account for the increase in the diffusion constant with S2 energy (i.e. more drifting electrons). Further study is needed to explain the energy dependence of $\sigma_0^2$. 

\section{Acknowledgments}
This work was supported by the US NSF (Grants PHY-0919363, PHY-1004072, PHY-1004054, PHY-1242585, PHY-1314483, PHY-1314507 and associated collaborative grants; grants PHY- 1211308 and PHY-1455351), the Italian Istituto Nazionale di Fisica Nucleare (INFN), the US DOE (Contract Nos. DE-FG02- 91ER40671 and DE-AC02-07CH11359), the Russian RSF (Grant No 16-12-10369), and the Polish NCN (Grant UMO-2014/15/B/ST2/02561). We thank the staff of the Fermilab Particle Physics, Scientific and Core Computing Divisions for their support. We acknowledge the financial support from the UnivEarthS Labex program of Sorbonne Paris Cit\'e (ANR-10- LABX-0023 and ANR-11- IDEX-0005-02), from S\~ao Paulo Research Foundation (FAPESP) grant (2016/09084-0), and from Foundation for Polish Science (grant No. TEAM/2016-2/17).

\bibliographystyle{ds}
\bibliography{ds}

\end{document}